\documentclass{emulateapj}
\newcommand{\ii}{$i'$}
\newcommand{\zz}{$z'$}
\newcommand{\uu}{$U$}
\newcommand{\bb}{$B$}
\newcommand{\vv}{$V$}

\newcommand{\zgal}{$z\!\simeq\!1$}
\newcommand{\sersic}{S\'{e}rsic}

\newcommand{\etal}{{et\thinspace al.}}

\newcommand{\Ho}{$H_{0}$}

\newcommand{\tabref}[1]{Table~\ref{#1}}
\newcommand{\figref}[1]{Figure~\ref{#1}}
\newcommand{\secref}[1]{\S~\ref{#1}}

\begin{document}

\title{Stellar Populations of Late-Type Bulges at \zgal\ in the HUDF}

\shorttitle{Late-type Bulges at \zgal\ in the HUDF}

\author{N.   P.  Hathi\altaffilmark{1,2},  I.  Ferreras\altaffilmark{3},
  A.     Pasquali\altaffilmark{4},   S.    Malhotra\altaffilmark{1,5},
  J.   E.    Rhoads\altaffilmark{1,5},   N.   Pirzkal\altaffilmark{6},
  R. A. Windhorst\altaffilmark{1,5} and C. Xu\altaffilmark{7}}

\email{Nimish.Hathi@ucr.edu}
\shortauthors{Hathi et al}

\altaffiltext{1}{Department  of  Physics,  Arizona  State  University, Tempe, AZ 85287-1504, USA}

\altaffiltext{2}{Department  of  Physics and Astronomy,  University of California, Riverside, CA 92521, USA}

\altaffiltext{3}{Mullard Space  Science Laboratory, University College
  London, Holmbury St Mary, Dorking, Surrey RH5 6NT, England}

\altaffiltext{4}{Max-Planck-Institut fuer  Astronomie, Koenigstuhl 17,
  D-69117 Heidelberg, Germany}

\altaffiltext{5}{School of Earth  and Space Exploration, Arizona State
  University, Tempe, AZ 85287-1404, USA}

\altaffiltext{6}{STScI, 3700 San Martin Drive, Baltimore, MD 21218, USA}

\altaffiltext{7}{Shanghai Institute of Technical Physics, Shanghai, China}


\begin{abstract}
  We  combine the  exceptional depth  of the  Hubble Ultra  Deep Field
  (HUDF)  images and  the  deep GRism  ACS  Program for  Extragalactic
  Science   (GRAPES)  grism  spectroscopy   to  explore   the  stellar
  populations  of  34  bulges   belonging  to  late-type  galaxies  at
  $0.8\!\le\!z\!\le\!1.3$.   The  sample  is  selected  based  on  the
  presence of  a noticeable 4000~\AA\  break in their  GRAPES spectra,
  and by  visual inspection  of the HUDF  images. The HUDF  images are
  used  to  measure  bulge   color  and  \sersic\  index.  The  narrow
  extraction of the GRAPES data around the galaxy center enables us to
  study  the  spectrum of  the  bulges  in  these late-type  galaxies,
  minimizing the contamination from the disk of the galaxy. We use the
  low resolution ($R\!\simeq\!50$)  spectral energy distribution (SED)
  around  the  4000~\AA\  break  to  estimate  redshifts  and  stellar
  ages.  The  SEDs  are  compared  with models  of  galactic  chemical
  evolution to determine the stellar mass, and to characterize the age
  distribution. We find that, (1)  the average age of late-type bulges
  in  our sample is  $\sim$1.3 Gyr  with stellar  masses in  the range
  10$^{6.5}$--10$^{10}$ M$_{\odot}$. (2)  Late-type bulges are younger
  than early-type  galaxies at similar  redshifts and lack a  trend of
  age with respect  to redshift, suggesting a more  extended period of
  star  formation.  (3) Bulges  and  inner  disks  in these  late-type
  galaxies show similar stellar  populations, and (4) late-type bulges
  are better  fitted by exponential surface  brightness profiles.  The
  overall picture emerging from the  GRAPES data is that, in late-type
  galaxies at  \zgal, bulges form through secular  evolution and disks
  via an inside-out process.
\end{abstract}

\keywords{galaxies: spiral --- galaxies: bulges --- galaxies:
formation --- galaxies: evolution}


\section{Introduction}\label{introduction}

There  are  currently  two  alternative  scenarios  to  explain  bulge
formation in galaxies.  First, semi-analytic models have traditionally
proposed  early  formation  from  mergers,  generating  a  scaled-down
version  of  an  elliptical galaxy  \citep[e.g.,][]{kauf93}.   Second,
dynamical instabilities can contribute to  the formation of a bulge in
a  primordial   disk  \citep{korm04}.   These   instabilities  can  be
triggered either  internally, or by  the accretion of  small satellite
galaxies  \citep{hern95},  and may  result  in  later  stages of  star
formation.  Hence,  the stellar  populations in galaxy  bulges provide
valuable constraints to distinguish between these two scenarios.

The  ability of  the \emph{Hubble  Space Telescope  (HST)}  to resolve
distant  galaxies enabled  the  study  of bulges  in  galaxies out  to
redshift    \zgal\   \citep{bouw99,abra99,elli01,mena01,koo05,maca08}.
Simple  phenomenological  models ---  such  as  the  one presented  in
\citet{bouw99} --- have, in the past, tried to determine whether bulge
formation  happens before  or after  the formation  of the  disk.  The
advantage  of the  lookback time  probed out  to \zgal\  allows  us to
quantify the  occurrence of merging vs.  secular  formation of bulges.
In  the  sample  presented  here  we  take  advantage  of  the  superb
capabilities  of the  Advanced  Camera for  Surveys  (ACS) to  extract
(slitless)  low resolution spectra  of bulges  from faint  galaxies at
these redshifts.

In their detailed review,  \citet{korm04} discuss two distinct type of
bulges, classical bulges --- i.e., merger-built with \sersic\ index $n
\gtrsim 2$ --- and pseudo bulges --- built out of disk material having
\sersic\ index  $n < 2$.   Early-type galaxies tend to  have classical
bulges,  while late-type  galaxies are  more likely  to host  a pseudo
bulge.   This  scenario  states  that early-  and  late-type  galaxies
generally form their bulges in  different ways. Many studies have been
done at low and high redshifts to investigate properties of bulges and
their    formation    histories.     Studies   on    local    galaxies
\citep{dejo96,cour96,thom06,caro07}  have  shown  through  colors  and
surface  brightness profiles  that later-type  galaxies in  the Hubble
sequence  have   more  bulges  best-fit  by   an  exponential  profile
(disk-like)  compared  to an  $r^{1/4}$  profile.  \citet{cour96}  and
\citet{dejo96} carried  out bulge-disk decompositions  for $\gtrsim$80
galaxies, and  found that 60--80\% of late-type  galaxies are best-fit
by  the double  exponential profiles.   More  recently, \citet{caro07}
used  \emph{HST} ACS  and the  Near Infrared  Camera and  Multi Object
Spectrometer (NICMOS)  multi-band imaging  to study the  structure and
the inner  optical and near-infrared colors  of local ($z\!\simeq\!0$)
bulges in a sample of nine late-type spirals.  Their analysis suggests
that half of the late-type  bulges in their sample must have developed
after the  formation of the  disk, while for  other half, the  bulk of
stellar  mass  was produced  at  earlier epochs  ---  as  is found  in
early-type  spheroids ---  and hence  must have  developed  before the
formation of  the disk.   \citet{thom06} analysed the  central stellar
populations of bulges  in spiral galaxies with Hubble  types Sa to Sbc
by  deriving luminosity-weighted  ages and  metallicities.   They find
that bulges are generally younger than early-type galaxies, because of
their  smaller  masses.   They  suggest  that  bulges,  like  low-mass
ellipticals, are  rejuvenated, but not by  secular evolution processes
involving disk material.
 
On the  theoretical side, semi-analytical and  $N$-body simulations of
galaxy formation have been mainly  based on two basic assumptions.  In
the  first  scenario, all  bulges  result  from  the merging  of  disk
galaxies \cite[e.g.,][]{kauf93}, whereas the second one is based on an
inside-out bulge  formation scenario \cite[e.g.,][]{vand98},  in which
baryonic  matter  of  a  protogalaxy  virializes  and  settles  in  an
inside-out process.  \citet{atha08} argues that in order to adequately
describe the formation and  evolution of disk-like bulges, simulations
should  include gas,  star formation  and feedback.   The  author also
states the  importance of cosmologically-motivated  initial conditions
in  the simulations, since  the properties  of pre-existing  disks may
influence the properties of the disk-like bulges.  When accounting for
these  effects, \citet{atha08} simulated  bulges that  show properties
similar to the observed disk-like bulges.

The     initial    studies    of     bulges    at     high    redshift
\citep{abra99,elli01,mena01} were done  by measuring optical colors in
galaxies to distinguish between bulge and disk colors.  \citet{elli01}
analyzed the internal optical colors of early-type and spiral galaxies
from  the  Hubble  Deep  Fields  \cite[HDF,][]{will96}  for  redshifts
$z\!\lesssim\!0.6$.   They  find  that  bulges  are  redder  than  the
surrounding  disks,  but  bluer  than  pure ellipticals  at  the  same
redshifts.  In  other work,  \citet{mena01} find strong  variations in
internal/central colors of more than  30\% of the faint spheroidals in
the HDF.  They do not  find such large variations in cluster galaxies,
and hence  estimate that  at \zgal, these  strong color  variations in
field bulges are due to more recent episodes of star-formation. Recent
studies \citep{koo05,maca08} have focused on bulge-disk decompositions
to  investigate the  colors and  radial profiles  of bulges  at \zgal.
\citet{koo05}  present a  candidate sample  of  luminous high-redshift
($0.73\!<\!z\!<\!1.04$) bulges ($I_{814}\!  <\!  23.1$ mag) within the
Groth  Strip Survey,  and find  that  majority of  luminous bulges  at
\zgal\  are very  red.   Their data  favors  an early  bulge-formation
scenario  in  which bulges  and  field  E-S0's  form prior  to  disks.
\citet{maca08}  study bulges  of spiral  galaxies within  the redshift
range  $0.1\!<\!z\!<\!1.2$  in the  Great  Observatories Origins  Deep
Survey (GOODS)  fields, and  find that bulges  of similar  mass follow
similar evolutionary patterns.

In this  paper, we use the  extraordinary imaging depth  of the Hubble
Ultra  Deep  Field  \citep[HUDF,][]{beck06}  and deep  slitless  grism
spectroscopy using  ACS from the  GRism ACS Program  for Extragalactic
Science (GRAPES) project  \citep[PI: S. Malhotra;][]{pirz04,malh05} to
explore the  stellar ages  of bulges in  late-type galaxies  at \zgal.
The exceptional  angular resolution and depth of  the GRAPES/HUDF data
combined with excellent grism sensitivity allows us to extract spectra
of the most central regions of faint galaxies at \zgal.

This paper  is organized as follows. The  \emph{HST} ACS observational
data,  sample selection  and  photometric properties  of the  selected
late-type    galaxies   are    presented    in   \secref{data}.     In
\secref{morpho},  we  discuss  morphological classification  of  these
galaxies.    The  non-parametric   CAS  measurements   are   shown  in
\secref{cas}, while we use GALFIT in \secref{galfit} to analyze galaxy
morphologies  in   two  dimensions.   We  study   the  bulges  stellar
populations via  Spectral Energy  Distribution (SED) fitting  of their
GRAPES  spectra  in \secref{models}.   Our  results  are discussed  in
\secref{results}  together  with  the   possible  biases  due  to  the
age-metallicity  relation,  and  our  conclusions  are  summarized  in
\secref{summary}.

Throughout this  paper we  refer to the  \emph{HST} ACS  F435W, F606W,
F775W,  and F850LP  filters as  the \bb-,  \vv-, \ii-,  and \zz-bands,
respectively.    We  adopt  a   Hubble  constant   \Ho=70~km  s$^{-1}$
Mpc$^{-1}$   and   a    flat   cosmology   with   $\Omega_m$=0.3   and
$\Omega_{\Lambda}$=0.7.   At redshift \zgal,  this cosmology  yields a
scale  of 1\arcsec  = 8.0  kpc. The  lookback time  is 7.7  Gyr  for a
universe that is 13.5~Gyr old.   Magnitudes are given in the AB system
\citep{oke83}.


\section{Observational Data and Sample Properties}\label{data}

\subsection{The \emph{HST}/ACS Data}

The HUDF is a 400 orbit survey of a $3.4'\times3.4'$ field carried out
with the  ACS in the \bb,  \vv, \ii\ and  \zz\ filters \citep{beck06}.
We have carried out deep  slitless spectroscopy of this field with the
ACS  grism as  a part  of  the GRAPES  project, which  was awarded  40
\emph{HST} orbits  during Cycle 12  (ID 9793; PI: S.   Malhotra).  The
grism observations were taken at five different orientations, in order
to minimize  the effects of contamination and  overlapping from nearby
objects.   The details  of the  observations, data  reduction  and the
final GRAPES  catalog are  described in \citet{pirz04},  who extracted
the grism spectra  for objects in the HUDF to  a limiting magnitude of
$z'_{\rm  AB}\!\simeq$27.5  mag. These  spectra  cover the  wavelength
range   between  6000  to   9500~\AA\  with   a  resolving   power  of
$R\!\simeq\!50\!-\!100$\footnote{Slitless   spectroscopy   produces  a
  variable spectral  resolution, depending on the size  of the object.
  The details concerning this  issue are described in \citet{pasq06a}}
and    are   characterized   by    a   net    significance   $N\!>\!5$
\citep[see][]{pirz04}.   We used  the multi-band  high-resolution HUDF
images to study  each object at \zgal\ in the ACS  band closest to the
\bb-band  rest-frame,  in  order   to  minimize  any  effects  of  the
morphological K-correction \citep[e.g.,][]{wind02}.

\subsection{Sample Selection and Properties}\label{sample}
 
\begin{figure}
\epsscale{0.85}
\plotone{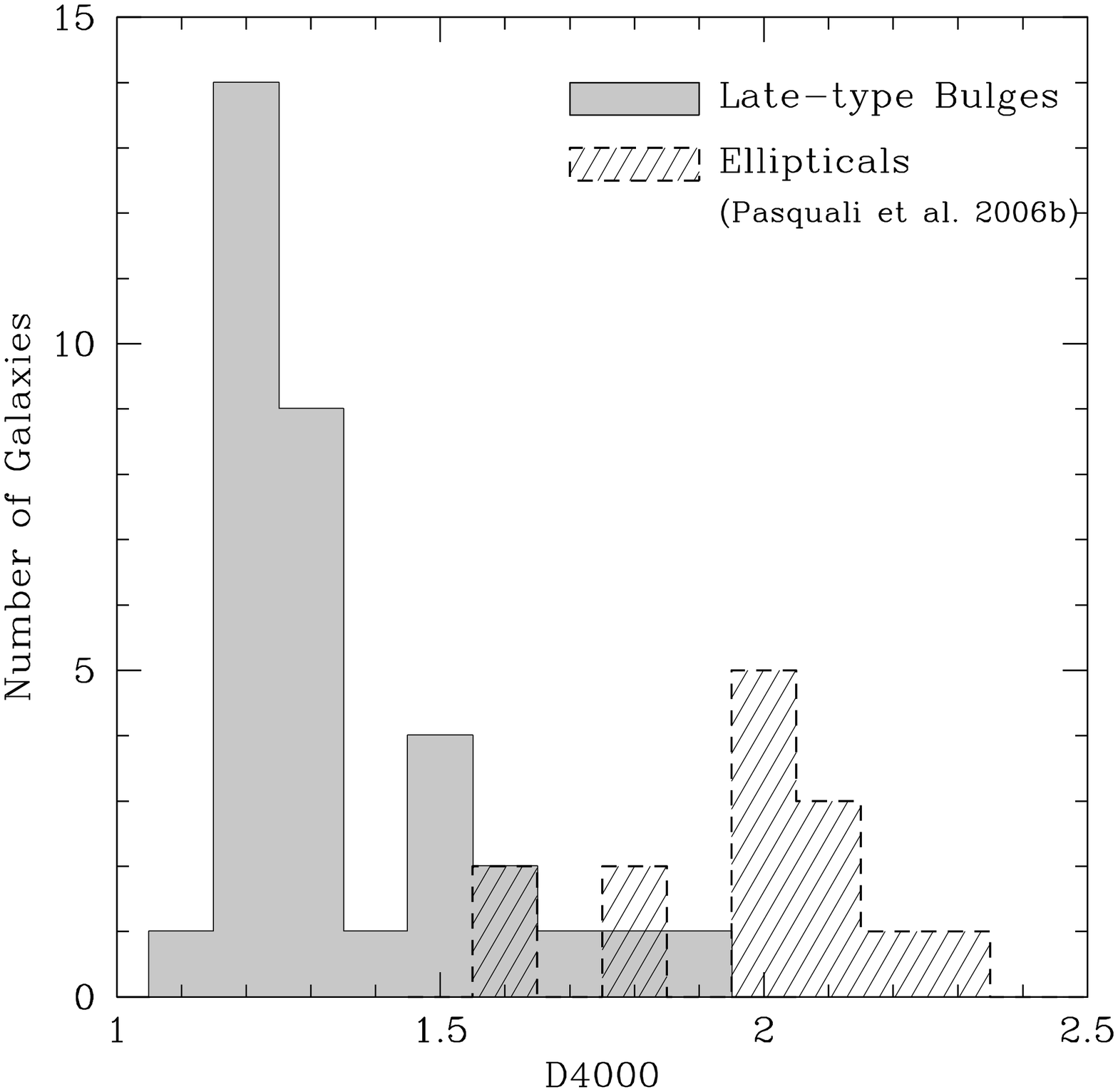}
\caption{Distribution     of    amplitude    of     4000~\AA\    break
  \citep[D4000;][]{balo99}  for  all  galaxies  in  our  sample  (grey
  histogram). D4000 was measured as  a ratio of average continuum flux
  redward and  blueward of  the 4000~\AA\ break  from 1D  GRAPES grism
  spectra.  The average value of the D4000 for our sample is $\sim$1.3
  and   for  comparison,   elliptical   galaxies  have   D4000~$>$~1.6
  \citep{kauf03,padm04}. We have also plotted D4000 (hashed histogram)
  for   GRAPES/HUDF  early-type   galaxies  ($0.6\!\le\!z\!\le\!1.1$).
  \citep{pasq06b}.}\label{fig1}
\end{figure}

\begin{figure}
\epsscale{0.85}
\plotone{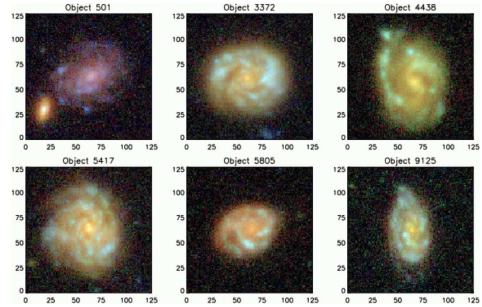}
\caption{Color  composite   images  of  a   representative  sample  of
  late-type spiral galaxies at \zgal.   Axes show size of the stamp in
  pixels (0.03\arcsec/pixel). Note that  all have small central bulges
  that are (in general) redder than their disks.}\label{fig2}
\end{figure}

The  ACS grism  sensitivity  peaks at  $\sim$8000~\AA,  and hence  the
GRAPES spectra are sensitive in identifying galaxies at \zgal\ through
their 4000~\AA\  breaks, and galaxies  at $z\!\simeq\!5\!-\!6$ through
their Lyman breaks.  We  selected objects with high signal-to-noise in
the  one-dimensional  (1D) GRAPES  grism  spectra  ($\sim$1500 in  the
HUDF).   All  these spectra  were  extracted  using narrow  extraction
windows  (5 pixels wide  --- the  pixel-scale in  the grism  images is
0.05\arcsec/pixel) around the center  of each galaxy.  We compared the
running average of flux in  10 data points between neighboring regions
on  each 1D  spectrum.   If the  difference  between two  neighboring,
average  flux values was  greater than  3$\sigma$, we  considered that
change  in  flux  level  as  a  `break'  in  the  spectrum.   A  large
($\sim$150) number of objects that show continuum breaks were selected
using this technique.  After  visual selection procedure, many objects
were   classified   and   studied   as   Lyman   break   galaxies   at
$z\!\gtrsim\!4.5$   \citep{malh05,hath08a},  or  as   late-type  stars
\citep{pirz05}, or  as ellipticals at  $z\!\simeq\!1$ \citep{pasq06b}.
During  this visual  classification, we  also found  that a  number of
late-type galaxies (mostly spirals) showed a prominent 4000~\AA\ break
(at observed $\sim$8000~\AA), which  is the major spectral feature due
to the  presence of  an old stellar  population.  This feature  was in
general  observed in  all grism  spectra obtained  for each  object at
different  position angles  and for  which spectral  contamination was
negligible.   We  selected  34  late-type  galaxies  (median  redshift
\zgal), to study the  properties of their central stellar populations.
We   measure  D4000  ---   the  amplitude   of  the   4000~\AA\  break
\citep[e.g.,][]{balo99} --- as the ratio of the average continuum flux
redward and blueward of the 4000~\AA\ break from the 1D grism spectra.
\figref{fig1} shows  the distribution of  D4000 for these  galaxies at
\zgal.   The  average D4000  ($\sim$1.3)  for  the selected  late-type
galaxies (grey  histogram) is  smaller than the  D4000 observed  for a
typical  elliptical galaxy \citep[$>1.6$,][]{kauf03,padm04}.   We also
show    D4000    (hashed    histogram)   for    elliptical    galaxies
($0.6\!\le\!z\!\le\!1.1$)  from \citet{pasq06b}.   \figref{fig2} shows
the  color-composite  images  of  6  representative  late-type  spiral
galaxies at \zgal\ in our  sample, which clearly demonstrates that all
have redder bulges in their centers.


\tabref{table1} shows the optical ($BVi'z'$) magnitudes, the HUDF IDs,
and  coordinates  for  each  selected  galaxy, as  obtained  from  the
published   HUDF   catalog  \citep{beck06}.    The   last  column   of
\tabref{table1}  gives  the   observed  (\vv--\zz)  colors  for  these
galaxies,  corresponding to  roughly rest-frame  (\uu--\bb)  colors at
\zgal.   \tabref{table2}  gives  all  available  redshifts  for  these
galaxies.  The  first 3 columns in \tabref{table2}  show the published
HUDF  IDs   and  coordinates,  while  the  fourth   column  shows  the
photometric  redshifts from  the  GRAPES spectro-photometric  redshift
catalog  \citep[for  description,  see][]{ryan07} or  the  GOODS-MUSIC
(MUltiwavelength  Southern Infrared  Catalog)  catalog \citep{graz06}.
The fifth column of  \tabref{table2} gives the spectroscopic redshifts
from VLT  \citep{graz06,vanz08}, when  available.  The last  column of
\tabref{table2}  gives  the  redshifts  from  GRAPES  SED  fitting  as
described  in  \secref{models}.  The  latter  are  the redshifts  used
throughout this  paper. Notice that  estimating the redshift  of faint
spectra  without  emission lines  can  be  challenging  even for  deep
surveys such as  VVDS \citep{lefe05}. Our slitless grism  data have an
optimal  spectral resolution  that maximises  the S/N  for populations
with a prominent 4000~\AA\ break. Comparisons of redshift estimates of
faint ($i_{\rm  AB}\!\simeq$22--24 mag) early-type  galaxies at \zgal\
from the Probing  Evolution And Reionization Spectroscopically (PEARS)
survey show that ACS slitless  grism data often fare {\sl better} than
publicly available  spectroscopic redshifts (Ferreras et  al. 2008, in
preparation).

\subsection{Observed Color Profiles}\label{colors}

\begin{figure}
\epsscale{0.85}
\plotone{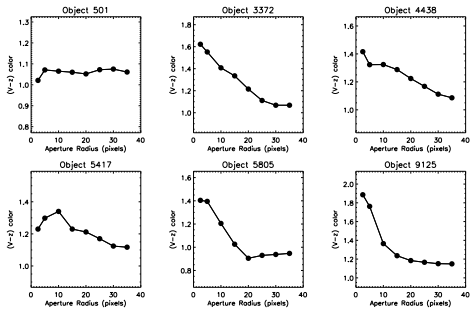}
\caption{Observed (\vv--\zz) colors measured using aperture magnitudes
  in  eight different  size apertures.   Here we  show colors  for six
  sample  galaxies.   These  are   the  same  six  galaxies  shown  in
  \figref{fig2}. Note  that with the  exception of object ID  501, all
  these galaxies become bluer from the inside outwards.}\label{fig3}
\end{figure}

We  used  the  Interactive  Data  Language  (IDL\footnote{IDL  Website
  http://www.ittvis.com/index.asp})                           procedure
\texttt{APER}\footnote{IDL    Astronomy    User's   Library    Website
  http://idlastro.gsfc.nasa.gov/homepage.html}   to  compute  aperture
photometry  using  several  aperture  radii.  We  chose  our  starting
aperture  radius to be  2.5 ACS  pixels ($\sim$0\farcs1),  because the
width  of  the  narrow  extraction window  \citep[see][for  extraction
details]{pirz04}  of our  GRAPES  spectra is  5  pixels.  We  measured
aperture magnitudes  for all  galaxies in our  sample in all  four ACS
bands ($BVi'z'$)  available from the HUDF with  aperture radii ranging
from 2.5  pixels to  35 pixels.  Using  these aperture  magnitudes, we
measured   the   (\vv--\zz)   color   profiles   for   our   galaxies.
\figref{fig3} shows these color  profiles for 6 sample galaxies.  This
figure show that the inner disk  in most of these galaxies is red, and
is dominated by the older  stellar population.  Our galaxies show that
for all apertures the (\vv--\zz) color is redder than 0.9, with redder
colors at smaller aperture sizes (the central bulge region).


\begin{figure}
\epsscale{0.85}
\plotone{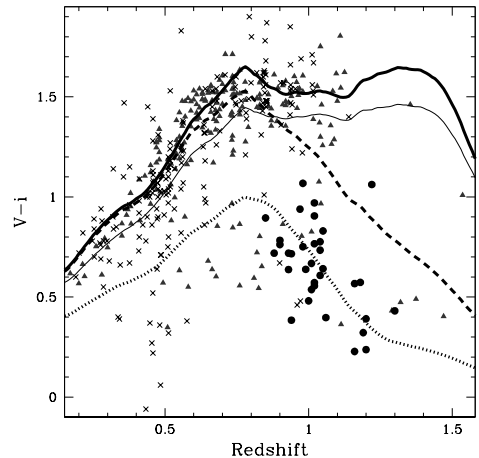}
\caption{Color vs. redshift diagram of bulges and spheroids. The black
  dots correspond  to our sample  of HUDF/GRAPES bulges.   The crosses
  are the  GOODS/HDF-N spheroids presented by  \citet{maca08}, and the
  triangles   are    the   GOODS/CDF-S   early-type    galaxies   from
  \citet{ferr05}.   The lines represent  the expected  color evolution
  for a set of  star-formation histories from the population synthesis
  models of \citet{bruz03}.  The thick lines track the color evolution
  of a stellar population  with solar metallicity and an exponentially
  decaying  star-formation  rate  starting  at  z$_F  =  5$,  with  an
  exponential decay  timescale of  0.5 (solid), 1  (dashed) and  8 Gyr
  (dotted). The thin  solid line shows the evolution  with a timescale
  of 0.5 Gyr at 1/3 of solar metallicity.}\label{fig4}
\end{figure}

\figref{fig4}  shows  the observed  (\vv--\ii)  color  as function  of
redshift for bulges  and spheroids.  The black dots  correspond to our
sample  of  HUDF/GRAPES  bulges.   The  crosses  are  the  GOODS/HDF-N
spheroids presented  by \citet{maca08},  and the filled  triangles are
the  GOODS/CDF-S early-type galaxies  from \citet{ferr05}.   The lines
represent the  expected color  evolution for a  set of  star formation
histories  using the  population synthesis  models  of \citet{bruz03}.
The thick  lines track  the color evolution  for a  stellar population
with  solar metallicity and  an exponentially  decaying star-formation
rate, starting  at z$_F = 5$,  with an exponential  decay timescale of
0.5 (solid), 1 (dashed) and 8 Gyr (dotted).  The thin solid line shows
the expected color evolution for a  decay timescale of 0.5 Gyr and 1/3
of solar  metallicity.  Most of  the bulges in early-type  galaxies in
GOODS/CDF-S  \citep{ferr05,maca08}  are  consistent with  short  decay
timescales,  while the  bulges in  our late-type  spiral  sample agree
better with a more extended star formation history.  This figure shows
that the bulges explored in  this paper belong to a similar population
as the  ``blue early-types''  presented in \citet{ferr05}.   This blue
population constitutes  $\sim$20\% of  the total sample  of early-type
systems visually  selected in  GOODS/CDF-S.  Because of  the excellent
photometric depth of  the HUDF, our sample extends  the bulge redshift
distribution of \citet{maca08} to $z\simeq1.3$.


\section{Morphological Properties}\label{morpho}

We performed  a morphological analysis  of the sample galaxies  in two
steps: (a) a non-parametric analysis of the distribution of the galaxy
light, using  the measures of asymmetry,  concentration and clumpiness
(or  smoothness)   to  confirm  our  visual  inspection;   (b)  a  two
dimensional (2D)  decomposition performed with  GALFIT \citep{peng02},
in  order to  quantify  the  galaxy morphology  and  in particular  to
extract their \sersic\ indices.

\subsection{CAS Measurements}\label{cas}

\begin{figure}
\epsscale{0.85}
\plotone{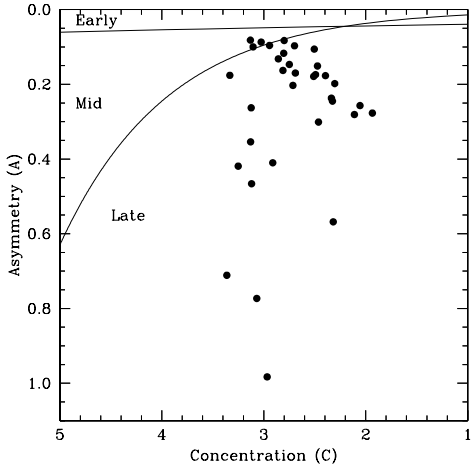}
\caption{Asymmetry and Concentration values for the selected late-type
  galaxies.   The  distinction  between  early-,  mid-  and  late-type
  galaxies  is from \citet{bers00}  and \citet{cons05}.   According to
  this classification  scheme, our sample -- selected  by the presence
  of a prominent 4000~\AA\  break and removing the spheroidal galaxies
  -- corresponds to late-type spirals.}\label{fig5}
\end{figure}

We use  the classical Concentration (C), Asymmetry  (A) and clumpinesS
(or smoothness)  --- the  CAS parameters \citep{cons00,cons03}  --- to
carry  out the  non-parametric  approach to  quantify morphology.   We
computed the C and A values for our galaxies following the definitions
and methods  as discussed in \citet{cons00}.   The Concentration index
correlates with  the \sersic\ index and the  Bulge-to-Disk ratio: high
C-values correspond to early-type morphology, while lower C-values are
suggestive of  a disk-dominated or later-type  and irregular galaxies.
Asymmetry can distinguish irregular galaxies or perturbed spirals from
relaxed  systems,   such  as  E/S0  and   normal  spirals.  Clumpiness
quantifies  the  degree  of  structure  on small  scale,  and  roughly
correlates  with   the  rate   of  star-formation.   We   derived  the
Concentration  and Asymmetry  indices using  the images  taken  in the
\zz-band,  which roughly  corresponds  to the  rest-frame \bb-band  at
\zgal.   Our measurements  of  C  and A  are  shown in  \figref{fig5},
together  with the  mean loci  for nearby  early-, mid-  and late-type
galaxies as derived by  \citet{bers00}.  A few perturbed galaxies show
higher asymmetry values.  \figref{fig5}  clearly shows that our sample
of galaxies  consists of mostly  late-type galaxies, and  confirms our
visual morphological classification of late-type galaxies with bulges.


\subsection{2D Galaxy Fitting using GALFIT}\label{galfit}

GALFIT \citep{peng02} is an  automated algorithm to extract structural
parameters from galaxy images by fitting/decomposing these with one or
more  analytic 2D  functions.  It  offers different  parametric models
(the ``Nuker''  law, the generalized  \sersic--de Vaucouleurs profile,
the exponential  disk, and Gaussian  or Moffat functions),  and allows
multi-component  fitting,  which is  useful  to measure  Bulge-to-Disk
(B/D) or Bulge-to-Total (B/T) light ratios.

\subsubsection{Thumbnail Image Extraction}\label{stamps}

The GALFIT  disk+bulge decompositions were performed  on thumbnail (or
``postage stamp'') images extracted  around the objects in our sample,
rather  than on  the  entire science  image  itself.  Three  thumbnail
images for each  object were extracted from the  original HUDF images.
All    thumbnail   images    are    201$\times$201   pixels    ($\sim$
6\farcs0$\times$6\farcs0) in  size. The first  thumbnail was extracted
from  the science image  itself.  The  second thumbnail  was extracted
from the comprehensive  segmentation image generated by \citet{coe06},
which was  used as  the ``bad pixel  map/mask'' image used  in GALFIT.
GALFIT uses this ``mask'' image so that all non-zero valued pixels are
ignored  in the fit.   Hence, the  extracted segmentation  stamps were
modified, so that only pixels  belonging to the galaxy had zero value,
while any  pixels belonging  to another object  are set to  a non-zero
value.  We tested this GALFIT decomposition with and without bad pixel
maps for  comparison, and obtained very similar  fitting results.  The
third thumbnail was extracted  from the drizzle-generated weight image
\citep{koek02}.   These weight  images  were modified  for GALFIT  (C.
Peng, private communication) as  follows. First, the science image was
smoothed  by  a  few  pixels  to   get  rid  of  some  of  the  random
pixel-to-pixel  variations.    Second,  a  variance   image  $S$,  was
calculated  using  $S=(1/wht) +  data/exptime$,  where,  $wht$ is  the
drizzle-generated  weight  image,  $data$  is  the  science  image  in
counts/sec  and $exptime$  is the  total exposure  time of  the image.
Finally, a sigma image is  generated using $\sigma = \sqrt{S}$.  These
modified weight images  were used as the input  ``sigma'' image (noise
maps) in the GALFIT, which is necessary for proper error-propagation.

\subsubsection{Sky Background}\label{sky}

The  drizzled  HUDF  images  are  sky  subtracted  and  therefore,  to
understand  the  effects of  sky-subtraction,  we  used the  following
procedure.   A careful  analysis of  the HUDF  sky-background  and its
corresponding uncertainties was performed by \citet{hath08b}.  We used
the sky-background values from  \citet{hath08b}, and allowed GALFIT to
either  vary this  sky-level during  the fitting  process, or  keep it
fixed.  For comparison, we  also used the sky-background measured from
each  individual object  stamp,  and repeated  the  process.  We  also
tested our  GALFIT decomposition by comparing the  results from GALFIT
with and  without the addition  of the sky-background.  We  found very
consistent  and  similar  fitting  parameters from  all  these  tests.
Therefore,   we  adopted   the  sky-background   levels   measured  by
\citet{hath08b} in our final GALFIT decomposition.

\subsubsection{Using GALFIT}

GALFIT  produces  model images  of  galaxies  based  on initial  input
parameters.   These images  are convolved  with the  ACS  Point Spread
Function (PSF)  image before comparing  with the actual  galaxy image.
Fitting  proceeds  iteratively until  convergence  is achieved,  which
normally occurs when the $\chi^2$ does not change by more than 5 parts
in   10$^4$    for   successive   five    iterations   \citep[see][for
details]{peng02}.

GALFIT requires initial guesses for the fitting parameters.  Following
\citet{dong06} and \citet{simi86}, we  used the output parameters from
the published HUDF SExtractor  catalogs \citep{beck06} as input values
for the magnitude, the half-light  radius, the position angle, and the
ellipticity of each object.  The initial value for the \sersic\ index,
$n$, was  taken to  be 1.5 \citep{coe06}.   Tests based on  an adopted
initial value of $n$=4 showed similar results.

\begin{figure}
\epsscale{0.85}
\plotone{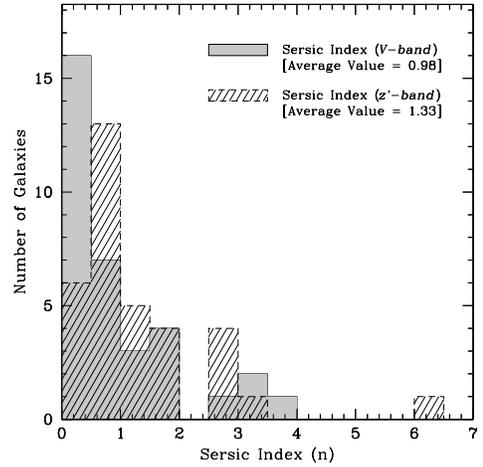}
\caption{The  distribution of  \sersic\ indices  ($n$) for  the galaxy
  retrieved from  GALFIT single component \sersic\ fit  to galaxies in
  \vv- and \zz-band.  The mean values  of $n$ is reported in top right
  corner. Using a two-sided K-S test on these distributions, we cannot
  reject the  hypothesis that the \vv- and  \zz-band distributions are
  drawn from the same population. }\label{fig6}
\end{figure}

All  GALFIT measurements  were  obtained from  the  \vv- and  \zz-band
images (approximately rest-frame  \uu- and \bb-band at $z\!\simeq\!1$,
respectively).   First,  we  fitted  only  an  one-component  \sersic\
profile  to our  galaxies to  improve  the initial  estimates for  the
\sersic\  index,  the  axis  ratio  and the  position  angle  of  each
galaxy. The distribution  of \sersic\ indices for our  galaxies in the
\vv- and \zz-band is shown  in \figref{fig6}.  Our measurements of the
\sersic\ index in \vv- and  \zz-band are comparable to \ii-band values
of the galaxies in \citet{coe06}.  \figref{fig6} confirms that most of
the galaxies  in our sample have  a \sersic\ index $n<2$  in both \vv-
and \zz-bands, which implies  that our galaxies are disk-dominated, as
\figref{fig2} and \figref{fig3} clearly suggest.


Next, we simultaneously fitted two components: a \sersic\ profile plus
an exponential disk profile, to  get better estimates of the bulge and
disk magnitudes, respectively.  For this simultaneous fit, we kept the
coordinates of  the galactic center (within  $\pm$1 pixel constraint),
the axial ratio and the  position angle fixed, while we allowed GALFIT
to  fit $m_{bulge}$,  $m_{disk}$, $R_e$,  $R_s$ and  the  the \sersic\
index ($n$).  For all galaxies  in our sample, we obtained better fits
for \emph{bulge} \sersic\ indices of $n \lesssim 1.0$.  We also tested
our runs by  fixing initial value of $n$=1 for  all galaxies and found
that GALFIT converged to similar  solutions in the end.  The bulge and
disk models obtained  from these best fits were  then used to estimate
the B/D  ratio.  We used an  aperture of 5 pixels  diameter to measure
the  B/D and  B/T ratios,  which used  the same  aperture size  as for
extracting the GRAPES  SEDs.  The majority of our  galaxies show a B/D
value $<$1  within this  aperture.  For larger  apertures encompassing
the galaxies' total light, B/D appears to be $<<$ 1, in agreement with
our galaxies being disk-dominated (i.e., late-type galaxies).

\section{Stellar Population Models - Bulge Properties}\label{models}

The GRAPES grism spectra were  taken at five different position angles
(PAs) to remove any contamination and overlap from nearby objects.  We
generated one final  spectrum for each galaxy by  combining all of the
GRAPES spectra obtained  at the 5 different PAs.   The combination was
performed  as  a  simple  averaging operation,  after  resampling  the
spectra onto a common wavelength grid.  Portions of spectra which were
contaminated  more than  25\%  \citep[see][for a  description]{pirz04}
were not  used, unless absolutely necessary.  The  Poisson errors were
propagated,  and the  standard deviation  of  the mean  between the  5
individual PAs was  computed.  The larger of either  the Poisson noise
or  the standard  deviation of  the mean  was used  in  the subsequent
analysis.   Our  goal is  to  fit  stellar  population models  to  the
age-sensitive 4000~\AA\ break observed  in the GRAPES spectra of these
galaxies.

\subsection{Star-Formation Histories (SFH)}

We fit our ACS grism spectra to a grid of models obtained by combining
the simple stellar populations  of \citet{bruz03}. A standard $\chi^2$
method is used.  We explore a  wide volume of parameter space in order
to infer robust constraints on  the possible ages and metallicities of
the stellar populations in the central bulges of these galaxies.  This
comparison requires  a careful process of degrading  the synthetic SED
(resolution $R\sim  2000$) to the (variable) resolution  of the GRAPES
spectra.  Special care must be taken with respect to the change of the
Line Spread Function  (LSF) with wavelength, which results  in both an
effective  degradation of  the spectral  resolution as  a  function of
wavelength, and  a different net  spectral resolution with  respect to
the size of the galaxy. After exploring a range of values $R\!=30-80$,
we find that an effective resolution of $R=50$ is suitable for all the
spectra  in our  sample.  The  ACS  grism spectral  resolution is  not
degenerate  with respect to  parameters describing  the star-formation
history, and mostly results in a global shift of the likelihood.

In order to determine the redshift as accurately as possible, we start
with some  guessing values obtained from three  sources: a photometric
redshift; a VLT  spectroscopic redshift --- where available  --- and a
redshift  estimate taken  from an  automated method  (as  discussed in
\secref{sample}) to  search for a prominent 4000~\AA\  break in GRAPES
data.  A small set of templates  at the GRAPES resolution were used to
determine the best redshift for each galaxy, using the guessing values
described  above  as  a   starting  point,  and  performing  a  simple
cross-correlation for  a range  of redshifts until  the best  match is
found. This method generates the redshifts used throughout this paper,
shown in \tabref{table2} as ``SED Fit'' (last column).

In order to make a robust  assessment of the ages and metallicities of
the  unresolved stellar  populations,  we use  two  different sets  of
models to describe  the build up of the  stellar component. The models
depend  on a  reduced  set  of parameters,  which  can characterize  a
star-formation history in a generic way.

 \noindent  {\bf Model  \#1  (EXP):} We  take  a simple  exponentially
 decaying star formation rate,  so that each star-formation history is
 well parametrized by  a formation epoch, which can  be described by a
 formation epoch  ($t(z_F)$); a star-formation  timescale ($\tau_\star
 =0.1\rightarrow  4$~Gyr); and  a  metallicity ($[m/H]=-1.5\rightarrow
 +0.3$), which  is kept fixed at  all times.  The  numbers in brackets
 give the range explored in the analysis of the model likelihood.  The
 range in formation epochs is chosen from $z_F=10$ to $t(z_F)=0.2$~Gyr
 (this range depends on the observed redshift of the galaxy).

 \noindent  {\bf Model  \#2 (CSP):}  We follow  a  consistent chemical
 enrichment code as described  in \citet{ferr00}. The model allows for
 gas infall  and outflow. The metallicity evolves  according to these
 parameters, using the stellar  yields from \citet{thie96} for massive
 stars ($>10M_\odot$),  and \citet{van97} for  intermediate mass stars.
 The free  parameters are the formation  epoch (same range  as the one
 chosen  for  Model  \#1),  the   timescale  for  the  infall  of  gas
 ($\tau_f=0.1\rightarrow 1$~Gyr),  and the fraction of  gas ejected in
 outflows   ($B_{\rm   OUT}=0\rightarrow   1$).   The   star-formation
 efficiency is kept  at a high value $C_{\rm  EFF}=20$ as expected for
 early-type populations \citep[see][]{ferr00}.

 For each of the two sets of  models we run a grid of SFHs, convolving
 simple  stellar populations  from the  models of  \citet{bruz03}. The
 grid  spans  $64\times  64\times  64$  SFHs  (note  that  three  free
 parameters are  chosen in each  set) over a  wide range of  values as
 shown above. Once the best fit  is obtained within the grid, we run a
 number  of  models with  random  values  of  the parameters  with  an
 accept/reject criterion  based on the likelihood --  analogous to the
 Metropolis  algorithm, e.g., \citet{saha03}.   The process  ends when
 10,000  models are  accepted.  The  total number  of  accepted models
 determine the  median and confidence  levels of the  parameters.  The
 distribution   of  reduced   $\chi^2$  values   has  an   average  of
 $\chi^2$=0.74  and  RMS $\sigma  (\chi^2)=0.35$.   The $\chi^2$  used
 throughout  includes a mild  Gaussian prior  on the  metallicity with
 average [m/H]$=-0.1$ and RMS $\sigma ([m/H])=0.5$.  This prior allows
 for  a  relatively  wide  range  of  average  metallicities,  and  is
 compatible  with   the  values   commonly  found  in   these  systems
 \citep[e.g.,][]{caro07}.  For the  interested reader,  we  include an
 appendix describing in  detail the effect of the  application of this
 prior. We find no significant  change in the estimates of the stellar
 age distribution with or without priors.

\begin{figure}
\epsscale{0.85}
\plotone{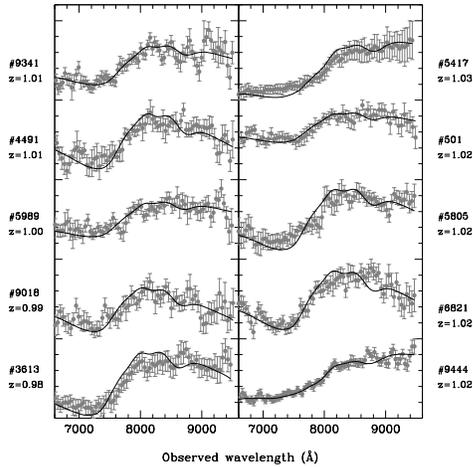}
\caption{Spectral energy  distributions of 10 galaxies  in our sample
  (each one  is labeled  with the HUDF  numbers and  model redshifts).
  The error bars are the observed ACS/G800L data and the lines are the
  best fits  according to the CSP  models (see text  for details). The
  distribution  of   reduced  $\chi^2$   values for all galaxies in our 
  sample has  an   average  of
  $\chi^2$=0.74 and RMS $\sigma (\chi^2)=0.35$.}\label{fig7}
\end{figure}

\begin{figure}
\epsscale{0.85}
\plotone{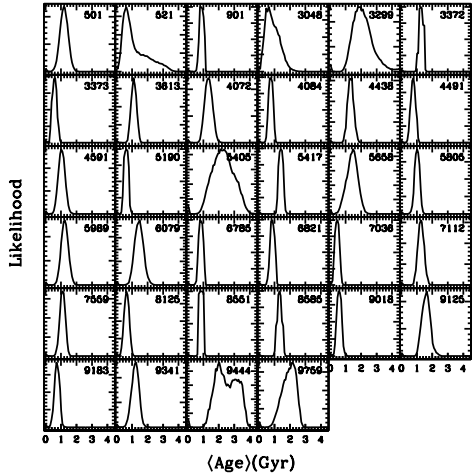}
\caption{The likelihood  distribution with respect to  average age for
  all 34 bulges.}\label{fig8}
\end{figure}

 \figref{fig7} shows the best fit models and the observed SEDs for 10
 galaxies in  our sample.  The error bars  represent the observations,
 and  the  solid  line  corresponds  to  the best  fits  for  the  CSP
 model. The  wavelength is  shown in the  observed frame,  whereas the
 wavelength range  chosen for all galaxies is  3800--5000~\AA\ {\sl in
   the  rest-frame}.   This  choice  ensures  a   consistency  in  the
 comparison of the stellar populations in our sample. The chosen range
 straddles  the  age-sensitive  4000~\AA\  break.  To  illustrate  the
 uncertainties  of estimating  stellar ages  within the  modelling, we
 show  in \figref{fig8}  the likelihood  distribution with  respect to
 average stellar age for each galaxy. 


\begin{figure}
\epsscale{0.85}
\plotone{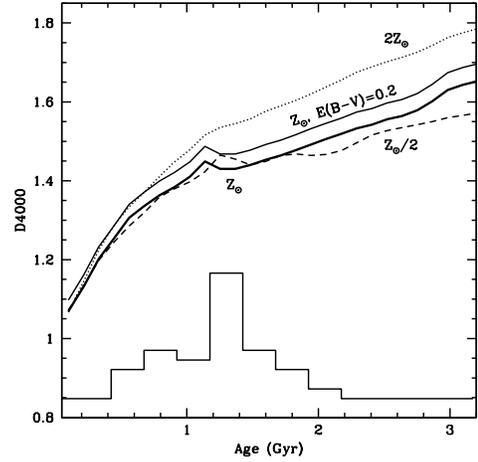}
\caption{The  4000~\AA\  break  amplitude,  D4000, as  a  function  of
  stellar  age.  The  distribution  of bulge  ages  is overplotted  as
  histogram,  which agrees  very well  with  the range  in D4000  (see
  \figref{fig1}).    Different  curves   show  three   simple  stellar
  populations model  with three different  metallicities, as labelled.
  The effect of  dust on a solar metallicity model is  shown as a thin
  solid line. The figure illustrates  that for the range of ages found
  in our  sample, the degeneracy  caused by dust and  metallicity will
  not  result in  a significant  change of  the average  stellar ages.
  Furthermore,  if the  bulges  were  dusty or  more  metal rich,  the
  resulting  populations would  become  even younger,  ruling out  the
  possibility  of bulges  as old  as early-type  galaxies at  the same
  redshift.}\label{fig9}
\end{figure}

 \figref{fig9}  shows  the  4000~\AA\  break amplitude,  D4000,  as  a
 function  of  stellar  age.    The  distribution  of  bulge  ages  is
 overplotted as a histogram, which  agrees very well with the observed
 range in  D4000 -- which  is 1.2 to  1.5, as shown  in \figref{fig1}.
 Different  curves show  three simple  stellar population  models with
 three   different  metallicities.    For  metallicity   around  solar
 (thick-solid  and dashed  lines), representative  of our  bulges, the
 variation in D4000  with age is very similar.   For solar metallicity
 with E(\bb--\vv)=0.2  dust reddening  (thin-solid line) and  for high
 metallicity  (dotted   line),  the  relation   is  somewhat  steeper.
 \figref{fig9} shows that for the  range in D4000 and metallicities of
 our sample, the effect of  metallicity on D4000 is small, hence D4000
 is a good age indicator for this sample.


\begin{figure}
\epsscale{0.85}
\plotone{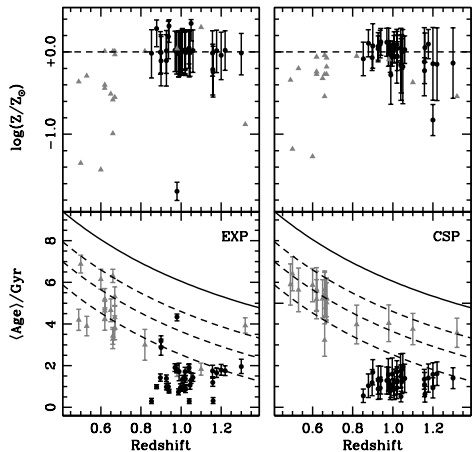}
\caption{Ages  and   metallicities  corresponding  to   the  best  fit
  according  to  a  simple  exponentially decaying  model  (EXP;  {\bf
    [Left]})  or  a consistent  chemical  enrichment  code (CSP;  {\bf
    [Right]}). The  filled circles  are the average  values of  age or
  metallicity   and  the  error   bars  represent   the  RMS   of  the
  distribution.  The solid lines in  the bottom panel track the age of
  the Universe  at a given  redshift for a concordance  cosmology. The
  dashed lines  -- from top to  bottom -- correspond to  the ages that
  {\sl simple  stellar populations} would have if  formed at redshifts
  $z_F=\{5,3,2\}$. The  grey triangles are  the GRAPES/HUDF early-type
  galaxies from \citet{pasq06b}, whose SEDs were analyzed the same way
  as in this paper.}\label{fig10}
\end{figure}

 \figref{fig10} shows the ages and  metallicities of the best SFHs for
 each bulge.  The average and  RMS scatter for age and metallicity are
 shown as  dots and error bars,  respectively.  For the  EXP models --
 which  have  zero  spread  in   metallicity  --  the  error  bars  in
 metallicity   represent   the    uncertainty   estimated   from   the
 likelihood. The solid lines in the lower panels correspond to the age
 of the Universe as a function  of redshift. The dashed lines show the
 age that  a {\sl  simple stellar population}  -- a population  with a
 single age -- would have if  formed at redshifts (from top to bottom)
 $z_F\!=\!\{5,3,2\}$.   The median value  of the  stellar ages  of our
 bulges is 1.3~Gyr. The CSP models treat chemical enrichment in a more
 consistent  way  than the  EXP  models  and  should therefore  better
 reflect the  true populations. We use mass-weighted  average ages for
 these stellar  populations from the SFHs because  they better reflect
 the  formation process  of bulges  (or galaxies  in general).  A very
 small amount  of young  stars -- something  that may not  reflect the
 true formation  process of the  bulge -- can  have a large  effect on
 luminosity-weighted ages.   By using composite models such  as EXP or
 CSP we  minimise this contamination  by using the  mass-weighted ages
 instead.


 We would emphasize here that it is the {\sl average} stellar age that
 can   be   reasonably   constrained   with  the   data.    Therefore,
 \figref{fig10}  does not  imply that  all stars  in these  bulges are
 $\sim$1.3 Gyr but many may be older.  To clarify, the formation epoch
 (characterized  by  a  formation  redshift)  is  the  age  when  star
 formation starts  in the model. Comparing observations  and models of
 unresolved stellar  populations can only give us  robust average ages
 (the first  order moment  of the age  distribution) and,  with higher
 uncertainty,  we  can  also   determine  the  ``width''  of  the  age
 distribution (the  second order moment).  We caution  the reader that
 the  actual parameters  used in  the modelling  (especially formation
 redshift)  constitute a  way to  characterise a  generic set  of star
 formation histories,  but the uncertainties in  these parameters will
 be larger than those in the  average age, which we consider to be the
 main physical property that can be extracted from the data.  We would
 also   clarify  here   that  we   have  used   `Age'   or  `\textless
 Age\textgreater' in  various figures showing  age distributions. When
 figures are based on single  stellar population models we use `Age',
 and  when figures are  based on  composite models  (EXP, CSP)  we use
 `\textless  Age\textgreater', as  they cannot  give a  single  age by
 definition.

\subsection{Bulge Mass Estimates}

The  photometry from  \tabref{table1}  can be  combined  with the  M/L
ratios obtained  from the best-fit  SFH to constrain the  stellar mass
(M$_s$) content of the bulges.  This M/L is derived from the composite
model  obtained  by  combining  the simple  stellar  populations  from
\citet{bruz03} using a \citet{chab03} Initial Mass Function (IMF).  If
we change  the IMF from \citet{chab03} to  \citet{salp55}, the stellar
mass will  increase by $\sim$0.3~dex  in $\log$(M$_s$),  which within
the  other errors  in data  and models,  does not  change  our overall
results.   The photometry  has to  be corrected  to take  into account
contamination from the disk, as we discuss in \secref{contam}.  We use
the B/D  ratio obtained from the GALFIT  (\secref{galfit}) to estimate
the  bulge fraction  of the  light in  the galaxy.   The  stellar mass
estimates  for  our  bulges  are  in  the  range  of  $6.5\!\leq\!\log
($M$_s/$M$_{\odot})\!\leq\!10.0$.

\begin{figure}
\epsscale{0.85}
\plotone{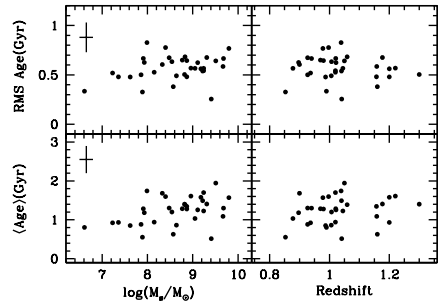}
\caption{The average age {\bf [Bottom]} and RMS scatter {\bf [Top]} of
  the age  distribution is  shown with respect  to stellar  mass ({\sl
    left}) and  redshift ({\sl right}). Typical error  bars are shown.
  The lookback time difference  between redshifts $z=0.8$ and $1.3$ is
  $1.9$~Gyr,  i.e., comparable to  the scatter  in {\sl  average ages},
  which  explains   the  lack   of  a  trend   of  average   age  with
  redshift.}\label{fig11}
\end{figure}

\figref{fig11}  shows  the  predicted  average  and  RMS  of  the  age
distribution  as a  function of  stellar mass  and redshift.  Over the
stellar masses and  redshifts probed in this sample  we find a similar
average age and scatter. This  similarity could be due to two possible
reasons.    First,   our   sample   spans  a   redshift   range   from
$z\!\simeq\!0.8$ to  $1.3$, corresponding to a  difference in lookback
time of $\sim 1.9$~Gyr. This  is comparable both to the uncertainty in
the  age estimate  and  to  the RMS  of  the distribution.   Secondly,
\citet{vanz06} found  Large Scale Structure (LSS) in  the CDF-S around
$z\!\simeq\!1.0$  from  VLT  spectroscopic  redshifts.   The  redshift
distribution  in the HUDF  (smaller field  in the  CDF-S) also  show a
strong peak around $z\!\simeq\!1.0$.  So it is possible that we may be
looking at a smaller subset of this LSS at $z\!\simeq\!1.0$.


\subsection{Disk Contamination in the GRAPES Grism Spectra}\label{contam}

The  best-fit stellar population  models to  the GRAPES  SEDs suggests
that the late-type bulges at \zgal\  are young, with an average age of
$\sim$1.3  Gyr. To  better understand  this result,  we first  need to
quantify the  effect of disk  contamination in our  measurements.  The
GRAPES SEDS are extracted  from an aperture of relatively narrow-width
aperture (5 pixels in diameter) around the center of each galaxy.  The
narrow extraction of  the grism spectra is dominated  by the bulge and
the inner  disk light; since  we use its  4000~\AA\ break to  date the
bulge, we need to investigate  the spectral contamination due to inner
disk  on the estimated  bulge age.   We perform  following photometric
tests to understand the effect of the inner disk on the bulge ages.

(1) We used  the disk and bulge light profiles  produced by GALFIT and
measured their  flux in a strip  5 pixel wide and  around their common
center,  to estimate  the disk  and bulge  light-fraction  within this
aperture.  We find that the light contributed by the disk to the total
flux in  this aperture can be  as high as  30\%. At the same  time, we
measured the disk  and bulge colors within the  same aperture, to find
that the disk and the bulge are similar within the photometric errors,
so that the  disk contamination in the bulge  spectrum is not expected
to dominate our estimate of the  bulge age (see Appendix B for further
discussion). This can  be already be seen in  \figref{fig3}, where the
bulge is in general 0.3--0.8 mag redder in (\vv--\zz) than the disk.

\begin{figure}
\epsscale{0.85}
\plotone{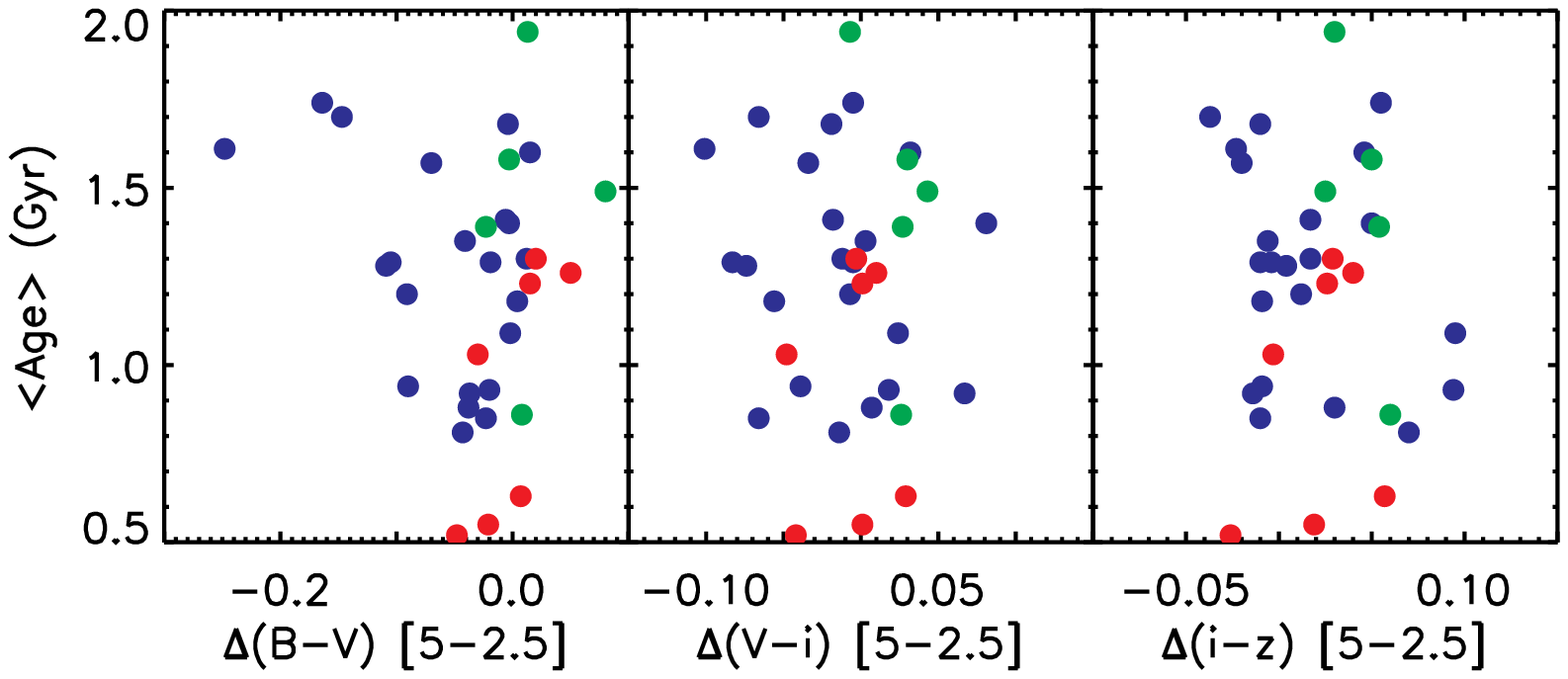}
\plotone{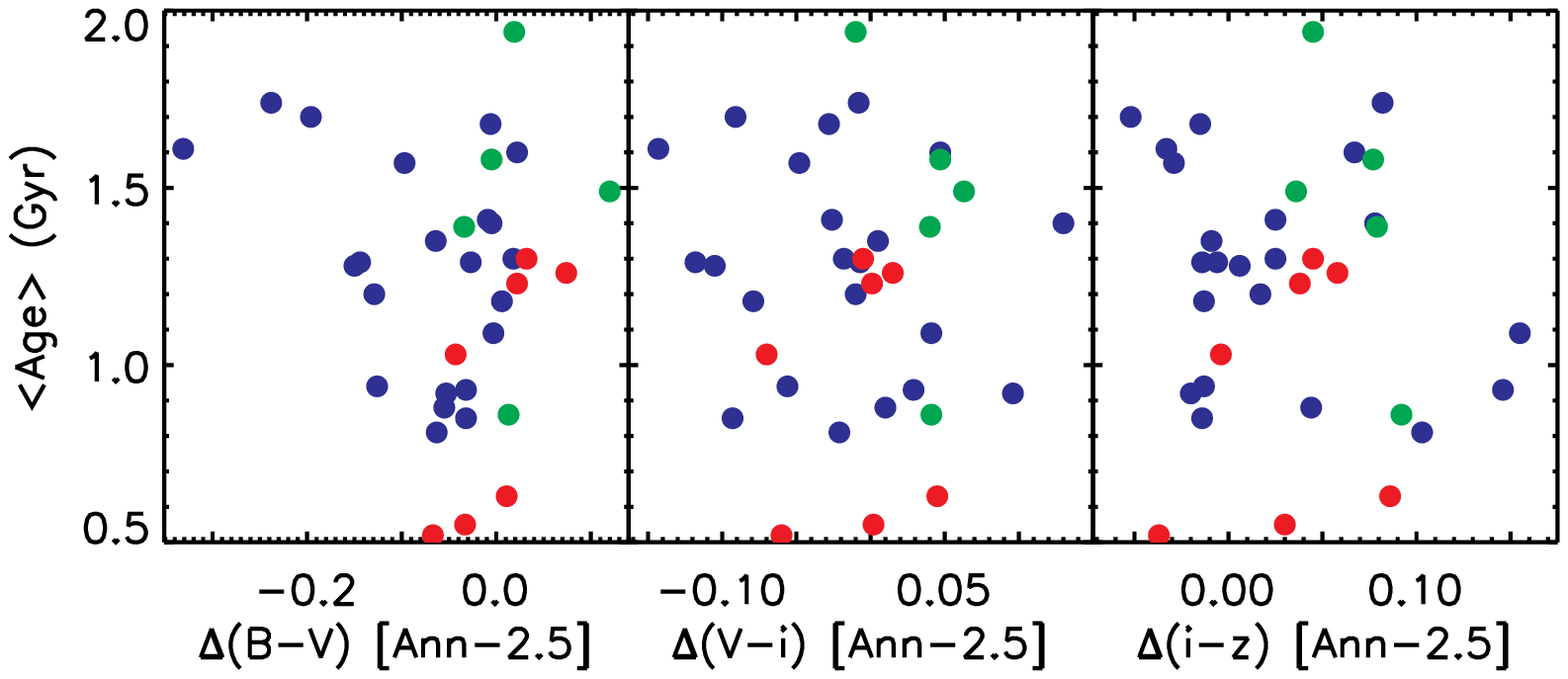}
\caption{Comparison  between aperture  colors and  the  best-fit bulge
  age. {\bf  [Top]} shows  the bulge  age as a  function of  the color
  difference  between  the  2.5  pixels  aperture  and  the  5  pixels
  aperture. {\bf  [Bottom]} shows the bulge  age as a  function of the
  color  difference between the  2.5 pixels  aperture and  the annulus
  defined by  the 2.5 and 5  pixels apertures.  A  blue colored circle
  stands  for   B/D$\le$0.5,  green  means   0.5$<$B/D$\le$1  and  red
  represents  B/D$>$1.   Here  B/D  is  measured in  the  \ii-band  as
  discussed  in  \secref{galfit}.  Both   panels  does  not  show  any
  correlation   among    the   age,   color    difference,   and   B/D
  ratio.}\label{fig12}
\end{figure}

\begin{figure}
\epsscale{0.85}
\plotone{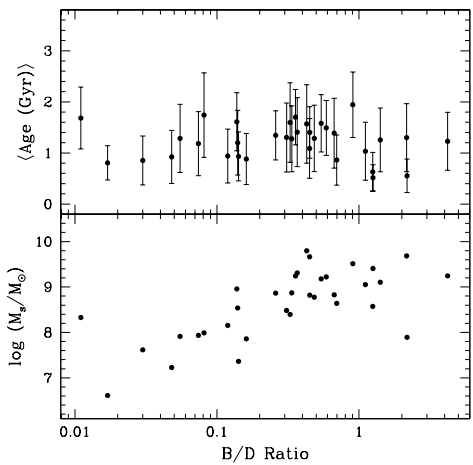}
\caption{The Bulge-to-Disk (B/D) ratio measured with GALFIT is compared
  with average age {\bf [Top]} and stellar mass {[\bf Bottom]}. Similar
  to the trend in \figref{fig11}, average age does not correlate with
  B/D either. However, the bottom panel suggests a correlation with
  stellar mass.}\label{fig13}
\end{figure}

(2)  We compared  the bulge  age derived  from the  stellar population
models with  the color difference  between two apertures.   We measure
the  color difference  between two  apertures  with 2.5  pixels and  5
pixels  radii,  equivalent to  the  narrow  and  wide GRAPES  spectral
extractions, respectively.  The top  panel of \figref{fig12} shows the
comparison between the color difference  and the bulge age. The points
on the  plot are color-coded  according to their B/D  ratios, measured
from GALFIT  as discussed in  \secref{galfit}.  Blue color  stands for
B/D$\le$0.5, green  means 0.5$<$B/D$\le$1 and  red represents B/D$>$1.
The top  panel of  the \figref{fig12} does  not show any  major trends
among age, color difference, and B/D ratio.  Secondly, we compared the
bulge age to the color  difference between the 2.5 pixels aperture and
the annulus  defined by  the 2.5 and  5 pixels apertures.   The bottom
panel of the \figref{fig12} shows  this comparison.  The points on the
plot  are color-coded according  to their  B/D ratios,  as in  the top
panel.  Like  the top panel,  the bottom panel of  \figref{fig12} does
not show  any major correlation  among age, color difference,  and B/D
ratio.  Finally, we  directly compared the B/D ratio  with the age and
mass  of  the  bulge.   The  top panel  of  \figref{fig13}  shows  the
comparison between  the B/D  ratios obtained from  the GALFIT  and the
bulge age from  the stellar population models for  all galaxies in the
sample. \figref{fig13}  does not show any correlation  between the age
and the B/D ratio. Similarly, the bottom panel of \figref{fig13} shows
at best a very mild correlation between the bulge stellar mass and the
B/D ratio.


(3) For a few sample galaxies, we extracted the spectrum of their disk
above and below the bulge aperture used to extract the bulge spectrum,
at  a distance  of approximately  10  pixels from  the galaxy  center.
Similarly to its bulge, the inner disk also exhibits a 4000~\AA\ break
in the spectrum  whose amplitude is only slightly  smaller (within few
percents) than the bulge.  This  test shows that both bulges and inner
disks are  equally red/old, as  expected from an  inside-out formation
scenario.   We also fitted  stellar population  models using  both the
bulge and disk spectra, and analyzed how the bulge-age and metallicity
change as a function of disk contamination, i.e., the fraction of disk
light  added to  the bulge  spectrum.  Our  simulations  (discussed in
Appendix  B) show  that  the determination  of  the bulge  age is  not
dominated  by  disk contamination.  Even  when  disk contamination  is
completely ignored  -- or  fully subtracted --  the bulge ages  do not
change much.

In summary, we  do not detect any significant  correlation between the
bulge  age, B/D  ratio  and  the aperture  color  difference. We  thus
conclude that  our estimate  of the bulge  ages is fairly  robust, and
that the younger  age of the sample bulges is likely  real and not due
to disk contamination.

\section{Discussion}\label{results}

The ages and  masses of late-type bulges are  estimated by fitting our
GRAPES SEDs  with stellar population models.  Our  analysis shows that
bulges in late-type  galaxies at higher redshift (\zgal)  appear to be
relatively young, with an  average age $\sim$1.3 Gyr ($6.5\!\leq\!\log
($M$_s/$M$_{\odot})\!\leq\!10.0$)  compared to early-type  galaxies at
the  same redshift.   This finding  appears to  be independent  of the
relative amount of disk-light present,  or the color of the underlying
disk.

\begin{figure}
\epsscale{0.85}
\plotone{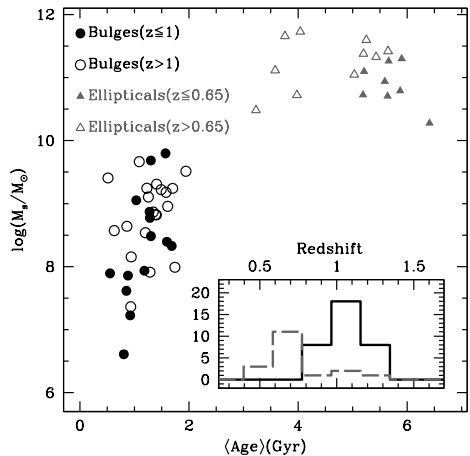}
\caption{Comparison  between the  ages of  the bulges  in  this sample
  (black  circles/lines)   and  early-type  galaxies   in  GRAPES/HUDF
  \citep[grey  triangles/dashed lines;  ][]{pasq06b}. The  inset shows
  the histogram of redshifts for both samples with ellipticals peaking
  around  $z\!\simeq\!0.65$. Both  samples are  split with  respect to
  redshift,  with  solid   symbols  representing  the  lower  redshift
  subsample.   There  is a  very  significant  difference between  the
  average age  of early-type galaxies and  galaxy bulges. Furthermore,
  the  age difference  is better  defined for  early-types, suggesting
  passive  evolution  for these  galaxies  and  a  more extended  star
  formation history for the bulges.}\label{fig14}
\end{figure}

\figref{fig14} shows the stellar masses of our bulges (filled and open
circles) compared with  the best-fit average ages from  the CSP models
discussed in \secref{models}. We also include the sample of early-type
galaxies  from GRAPES/HUDF  \citep[grey  triangles;][]{pasq06b}, whose
GRAPES spectra were analyzed in a similar way.  The early-types sample
covers a wider range of redshifts ($0.5 \le z \le 1.1$).  Hence, for a
proper comparison, we divide  both bulges and early-type galaxies with
respect to redshift roughly about  the median value for each subsample
$z\sim  1$ for  bulges and  $z\sim  0.65$ for  early-types. The  solid
(hollow)  symbols  correspond  to  the lower  (higher)  redshift  bin,
respectively.  The bulges  in our late-type spirals span  a much lower
range of ages, and, obviously,  have lower stellar masses, compared to
those of early-type galaxies.


\citet{elme05}  have  classified $\sim$900  galaxies  (larger than  10
pixels or  0.3\arcsec) in the  HUDF according to morphology  and their
photometric properties.   They find 269  spiral galaxies in  the HUDF.
Using the  \citet{elme05} morphological classifications,  and accurate
spectro-photometric  redshifts from  \citet{ryan07}, we  estimate that
the results  in this  paper represent approximately  $\sim$40--50\% of
the  total   late-type/spirals  HUDF  galaxy   population  within  the
magnitude and redshift range used in this paper.

Our  analysis  of the  central  and the  inner  disk  colors of  these
galaxies (\figref{fig3}) and their  grism spectra (as discussed in the
Appendix B)  shows that the inner  disk and the  bulge components have
similar colors, and that the bulge ages are not significantly affected
by the light  (and stellar populations) of the  underlying disk.  This
result is consistent with the idea  that the inner disk of galaxies in
general    has    similar    colors    and   age    as    the    bulge
\citep[e.g.,][]{pele96}.   The effect  of dust  on  these measurements
should  not be  significant, since  our bulge  ages are  based  on the
amplitude of the 4000~\AA\ observed in the GRAPES grism spectra, which
is mostly  sensitive to  age and  has a weak  dependence on  dust (see
\figref{fig9}). Also, \citet{maca04} argue  that dust is generally not
a significant contributor  to galaxy colors in low-mass/low-luminosity
spiral  galaxies, but  is  likely important  in more  massive/brighter
galaxies. On  the other hand,  even if it  plays an important  role in
this analysis, the inclusion of  dust will make our ages even younger,
and  our result  that  bulges and  inner-disks  have similar  dominant
stellar population with an average age of $\sim$1.3 Gyr should then be
viewed as an upper limit.

We  performed   GALFIT  decomposition   on  the  sample   galaxies  by
simultaneously fitting the bulge to a \sersic\ profile and the disk to
an exponential  profile.  For  all bulges in  our sample,  we obtained
better  fits using \sersic\  indices of  $n \lesssim  1.0$. Therefore,
these  bulges  are  disk-like  \citep{korm04,atha05} and  have  radial
surface brightness  profiles similar  to disks.  Similar  analyses for
local spirals by \citet{dejo96} and \citet{cour96} have shown that the
majority of bulges in late-type galaxies are better fit by exponential
profile.  Our results show that  a similar trend also exists at \zgal.
The similarities  we find in  the bulge and the  inner-disk properties
(4000~\AA\  break, colors  and profiles) could  imply that  these less
massive,  younger  bulges at  \zgal\  grow  through secular  evolution
processes \citep{korm04}.  At \zgal, it is possible that we are seeing
these  galaxies still  forming, and  these ``disk-like''  bulges might
grow from disk material or minor mergers to become more massive bulges
observed at present day.  Disk-like pseudo-bulges can also grow by gas
inflow  and  star-formation.   Bars  can  drive  central  gas  inflows
\citep{shet05}, and  therefore, there  could be a  correlation between
these disky bulges and central bars.  \citet{shet08} find that the bar
fraction  in   very  massive,   luminous  spirals  is   constant  from
$z\!\simeq\!0$  to  $z\!\simeq\!0.84$,   whereas  for  low-mass,  blue
spirals it declines significantly with redshift to about $\sim$20\% at
$z\sim0.84$,  indicating   that  some  bars  do   form  early  enough.
\citet{elme05} has morphologically  classified few ($\sim$10\%) of our
sample  galaxies as  barred galaxies,  so  it will  be interesting  to
investigate these  late-type galaxies in future  studies to understand
this relation.

Aperture   color   analysis    by   \citet{elli01}   for   bulges   at
$z\!\lesssim\!0.6$ in  early-type and  spiral galaxies with  $I_{AB} <
24$  mag  found  that  their   central  colors  are  redder  than  the
surrounding outer disk colors, but that these central colors are bluer
than those  of pure  ellipticals at the  same redshifts.  As  shown in
\figref{fig10},  our results  agree with  \citet{elli01}. This  is not
perhaps surprising,  since we also  select our sample based  on galaxy
total-magnitudes,  with no  constraints  on its  bulge magnitude.   In
comparison,   \citet{koo05}  select  their   sample  based   on  bulge
luminosity,    and   they    find    that   luminous,    high-redshift
($0.73\!<\!z\!<\!1.04$) bulges  ($I_{AB} <  24$ mag) within  the Groth
Strip Survey  are very  red/old. They clearly  show that if  the bulge
sample  is luminous,  then all  bulges are  equally red  and  old.  In
contrast, we  show that  if the bulge  sample is selected  without any
constraint on  the bulge magnitude, then late-type  bulges are younger
than bulges in early-type galaxies at similar redshifts.

Galaxy colors  and structural properties show  a bimodal distribution,
separating into a red  sequence, populated by early-type galaxies, and
a     blue    ``cloud'',     populated    by     late-type    galaxies
\citep{balo04,driv06}. Whether a galaxy resides in a red sequence or a
blue  cloud  is  also  related  to  the type  of  bulge  in  a  galaxy
\citep{dror07}.  \figref{fig4}  shows this bimodal  distribution.  Our
late-type galaxies with pseudo bulges  lie in the bluer cloud compared
to early-type galaxies that lie on the red color sequence.  This shows
that the  processes involved in  the formation of galactic  bulges and
their  host galaxies  are  very similar.   Observations indicate  that
these formation  mechanisms depend strongly  on the bulge (as  well as
galaxy)  mass, and that  they were  active at  $z\!\simeq\!1.3$.  This
evidence  is  strengthened  by   the  results  of  \citet{thom06}  and
\citet{maca08}, who  find that bulges  of similar mass have  a similar
evolutionary  path. Possibly  because of  cosmic variance,  we  do not
detect early-type spirals at \zgal\ in the HUDF.  Comparing our sample
of late-type  bulges with the  massive early-type galaxies  at similar
redshifts  studied by  \citet{pasq06b},  we confirm  the existence  of
different  evolutionary patterns  for bulges  in early-  and late-type
galaxies (see \figref{fig4}).

Our analysis of  the deepest optical survey, the  HUDF, along with the
deep  unique   ACS  grism  spectroscopy  provides   the  best  spatial
resolution at \zgal,  and yields more detailed insight  in the process
of     galaxy    formation.      Massive    and     luminous    bulges
\citep[e.g.,][]{koo05}, which mostly reside in early-type galaxies and
in   earlier-type   spiral   galaxies,   are   old   and   formed   at
$z\!\gtrsim\!2$.   The secular  evolution suggested  by \citet{korm04}
does not play any role in their formation.  Our ACS grism study of the
HUDF  here  has  shown   that  lower-mass  bulges,  which  are  mostly
associated with  later-type galaxies, are on average  younger than one
would  expect by  letting their  stellar populations  passively evolve
since their  formation redshift $z\!\sim\!2$. This  result would point
to secular evolution  as a likely mechanism to  support prolonged star
formation   in  low-mass   bulges,  which   is  very   different  from
\citet{pasq06b}  ellipticals,  for  which  a quick  SFH  plus  passive
evolution can explain their observed SEDs.

\section{Summary}\label{summary}

We estimated  the stellar  ages and masses  of 34 bulges  of late-type
galaxies from  the HUDF  by fitting stellar  population models  to the
4000~\AA\ break observed in deep ACS GRAPES grism spectra.  This study
takes advantage of the exceptional angular resolution and depth of the
GRAPES/HUDF  data,  which  allow   us  to  identify  the  small  bulge
components of  this sample  of galaxies at  \zgal, \emph{both}  on the
direct   \emph{and}  on  the   grism  images,   and  to   extract  its
corresponding   spectrum,   a   method   currently   unfeasible   with
ground-based spectroscopy.  We find that bulges  in late-type galaxies
at higher redshift (\zgal) appear to be significantly younger (with an
average  age of  $\sim$1.3 Gyr)  than early-type  galaxies  at similar
redshifts.  This  finding is robust  against the amount  of disk-light
that may contaminate in part the  bulge spectrum.  The lack of a trend
between average  age and  redshift (see \figref{fig14})  suggests star
formation in  bulges is  extended over much  longer times  compared to
early-type galaxies.  Our results support the  scenario where low-mass
late-type bulges form through secular evolution processes.

\acknowledgments 
This work was supported by \emph{HST} grants GO 9793 and GO 10530 from
the Space Telescope Science Institute, which is operated by AURA under
NASA contract NAS5-26555.   We would like to thank  Seth Cohen for his
help generating color images and  Chien Peng for his prompt replies on
our GALFIT  questions.  NPH would like to  thank Graduate Professional
Student  Association  (GPSA) at  Arizona  State  University for  their
conference travel grant  to present this work at  the AAS meeting.  IF
thanks the  School of  Earth and Space  Exploration (SESE)  at Arizona
State  University  for hospitality  and  financial  support, and  also
acknowledges  support  from  the  Nuffield Foundation.  We  thank  the
referee for helpful comments and suggestions that improved the paper.

Facilities: \facility{HST(ACS)}

\appendix

\section{Effect of Metallicity Priors on Bulge Ages}

\begin{figure}
\epsscale{0.6}
\plotone{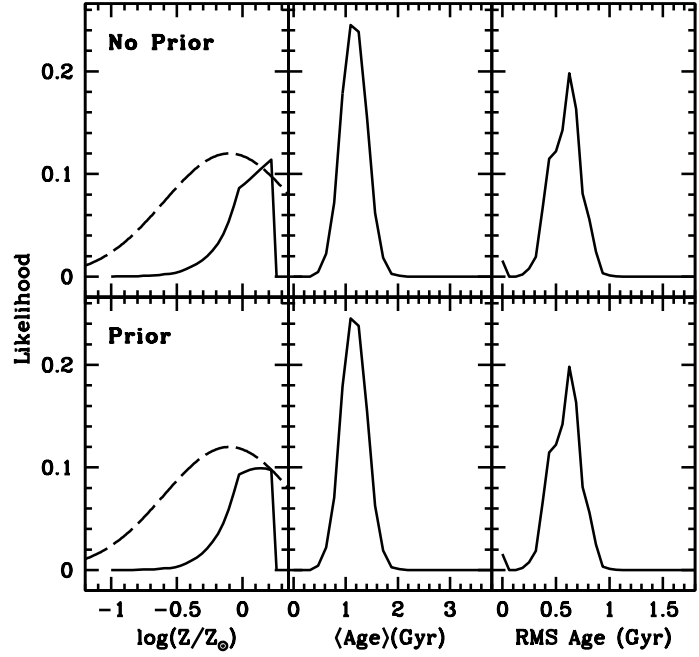}
\caption{One typical example (ID  501) of the likelihoods obtained for
  metallicity, average age  and RMS of the age  distribution for a set
  of EXP models with and  without metallicity prior. This figure shows
  that without the prior we  get a slightly higher metallicity but the
  average  stellar  age  and  scatter  remains  very  similar  to  the
  modelling with a prior.}\label{fig15}
\end{figure}

The modelling  of stellar  populations in this  paper includes  a mild
prior on the distribution of bulge metallicities. For each choice of a
star formation  history (characterized by the  parameters discussed in
\secref{models})  we   multiply  the  likelihood   from  the  $\chi^2$
distribution by a Gaussian prior with mean $\log (Z/Z_\odot)=-0.1$ and
RMS  $\sigma  (\log Z/Z_\odot)=0.5$.   The  effect  of  this prior  is
illustrated  in \figref{fig15}  for one  of the  bulges (ID  501). The
marginalised distribution  in metallicity,  average age and  age width
(RMS) is  shown for the modelling  without ({\sl top})  and with ({\sl
  bottom}) the metallicity prior. The  Gaussian prior is also shown as
a dashed  line in the leftmost  panels. The main effect  of this prior
(and the  motivation to our inclusion  of this term in  this paper) is
that  the metallicity  likelihood has  a monotonic  increase  for high
metallicities,  above solar. Hence,  we decided  to reduce  weight for
those  models with  very high  metallicities by  the inclusion  of the
prior.   This  decision  is  justified  by the  fact  that  population
synthesis models are not as  accurate at super solar metallicities due
to the  lack of  proper calibrators \citep{bruz03}.   Furthermore, the
range  of  metallicities  for  which  the  prior  has  a  mild  effect
corresponds  to  the  range  commonly  found  in  these  systems  from
ground-based                 spectroscopic                observations
\citep[e.g.,][]{shap05,schi06,liu08}.  The effect  of the prior on the
age  distributions  is  negligible  (middle and  rightmost  panels  of
\figref{fig15}).  For  those reasons we  believe the inclusion  of the
prior does  not bias  in a significant  way our  conclusions regarding
stellar ages.


\begin{figure}
\epsscale{0.6}
\plotone{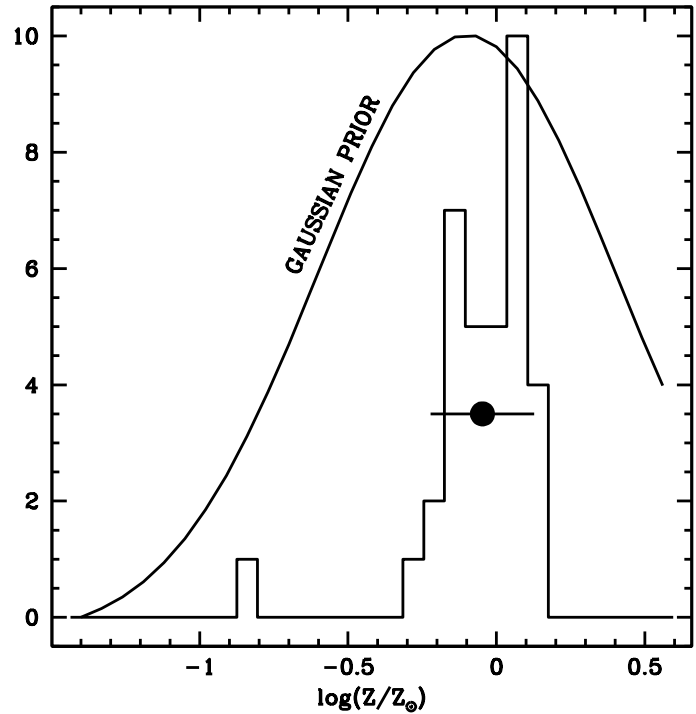}
\caption{The  distribution of bulge  metallicities estimated  from the
  stellar population  models. The data  point shows the  average value
  and RMS  scatter of  the distribution. We  overplot a  mild Gaussian
  prior used on the  metallicity with average $\log (Z/Z_\odot) =-0.1$
  and RMS=$0.5$~dex.   This prior allows  for a wide range  of average
  metallicities, and is compatible  with the values obtained at \zgal\
  from         ground-based         spectroscopic         observations
  \citep[e.g.,][]{shap05,schi06,liu08}.}\label{fig16}
\end{figure}

But we can also show that the prior should not have a strong effect on
the distribution of metallicities either.  \figref{fig16} compares the
predicted metallicities of our  sample (histogram).  The filled circle
and  error  bar  represents  the  mean  and  RMS  of  the  metallicity
distribution. In  comparison, the Gaussian prior (solid  line) is much
wider.
Only when the prior  and the predicted values have the same
distribution one could suspect of strong biasing.


\begin{figure}
\epsscale{0.6}
\plotone{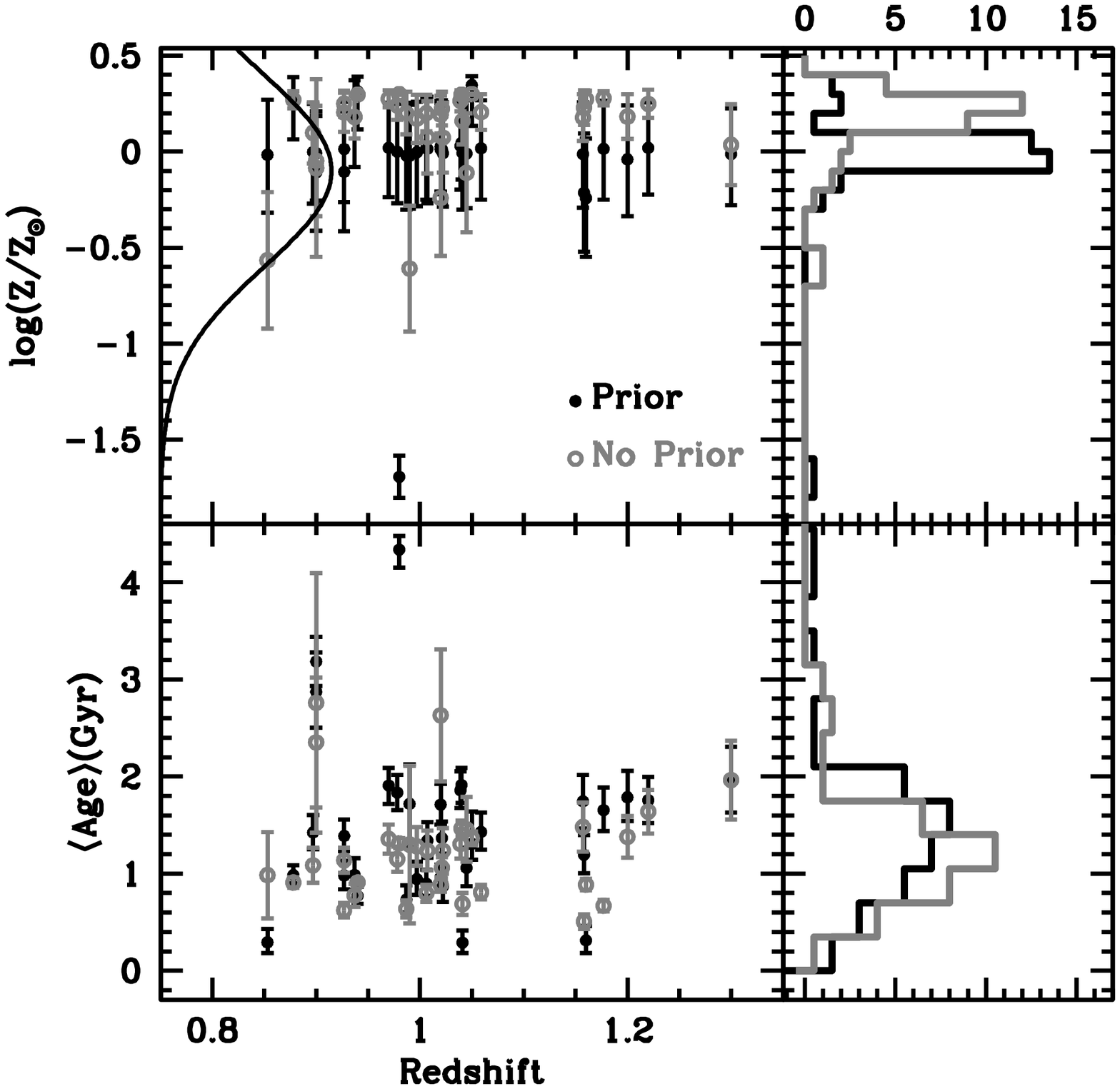}
\caption{Comparison  of bulge age  distribution and  bulge metallicity
  for  EXP  models with  and  without  a  metallicity prior.  The  age
  histogram  does not  show  any appreciable  change. The  metallicity
  histogram   without  a   prior  (grey)   peak  at   slightly  higher
  metallicities -- as illustrated in \figref{fig15} -- compared to the
  metallicity  histogram (black)  with prior,  but it  is nevertheless
  compatible  given  the typical  uncertainty  (of  order 0.3~dex)  in
  estimates     of     metallicity     from     unresolved     stellar
  populations.}\label{fig17}
\end{figure}

Finally, in order to quantify  the effect of metallicity priors on the
predicted ages and  metallicities for our sample, we  show a comparison
for EXP  models with  (black dots) and  without priors (grey  dots) in
\figref{fig17}. The rightmost panels  show histograms of the predicted
ages  and  metallicities.   As  expected, the  metallicities  obtained
without a prior are higher  compared to models with no prior. However,
taken  into  account  reasonable  uncertainties  for  the  metallicity
estimates extracted  from unresolved  spectra (of order  0.3~dex), one
can say that  our methodology is acceptable within  the expected error
bars.


\section{Effect of Disk Contamination on Bulge Ages}

\begin{figure}
\epsscale{0.6}
\plotone{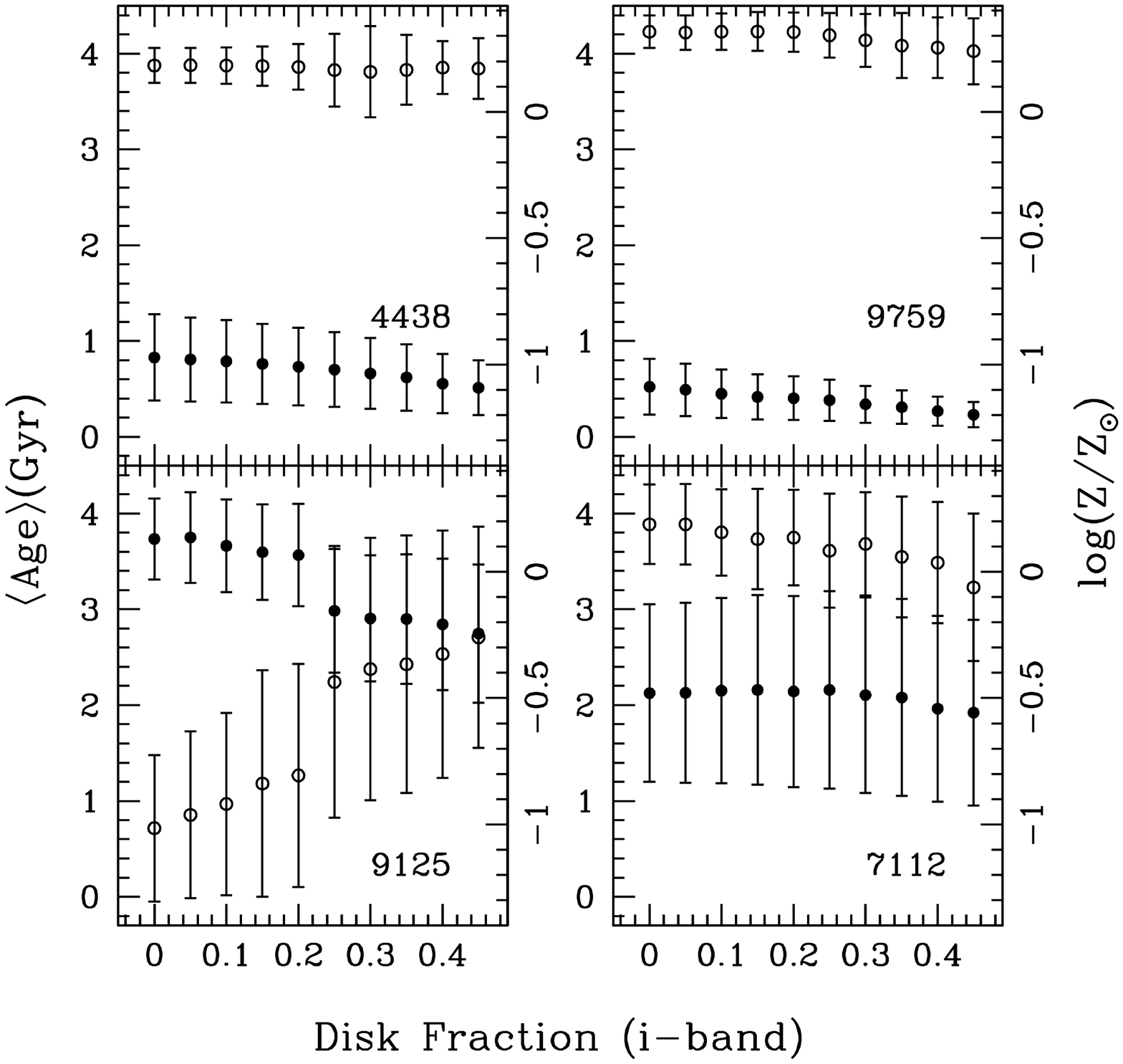}
\caption{Effect of  disk contamination on  the predictions of  age and
  metallicity.  The solid (empty)  dots show the estimated average age
  (metallicity)  for a  grid of  exponentially-decaying star-formation
  histories.  The 95\% confidence levels  are shown as error bars. The
  result is presented as a  function of the contamination of the disk,
  as measured in the F775W band (see text for details).}\label{fig18}
\end{figure}

In order  to assess the  effect of disk  contamination on the  age and
metallicity estimates,  we performed an  extraction of the SED  of the
disk component for a small sub-sample of galaxies. This corresponds to
the stacking of two strips  (5 pixels wide) at an approximate distance
of  10 pixels  on  either side  of  the center  of  the galaxy.   When
comparing the observed spectra with synthetic models, we assume that a
fraction of  the flux in each  galaxy comes from the  disk. Hence, for
each choice of parameters, a synthetic SED is obtained, and a fraction
$f_D$ of the  disk SED is added to this  spectrum, before performing a
maximum  likelihood  analysis.   This  fraction  is  measured  in  the
\ii-band  (F775W).   \figref{fig18}   shows  the  resulting  ages  and
metallicities for four galaxies from  our sample as a function of disk
contamination.  The  95\% confidence levels  are shown as  error bars.
The ages  and metallicities are shown  in the same plot,  as solid and
open  dots,  respectively.   One  can  see that  the  effect  of  disk
contamination  is small,  as expected  from the  weak  color gradients
found in the images.  This result is consistent with the idea that the
inner disk of galaxies is as old as the bulge \citep[e.g.,][]{pele96}.


\clearpage


\begin{deluxetable}{cccccccc}




\tablecaption{Our Sample - HUDF IDs, Coordinates, Optical Magnitudes and Colors \label{table1}}

\tablenum{1}

\tablehead{\colhead{HUDF} & \colhead{RA (J2000)} & \colhead{DEC (J2000)} & \colhead{$B$} & \colhead{$V$} & \colhead{$i'$} & \colhead{$z'$} & \colhead{(\vv--\zz)$^{\dagger}$} \\ 
\colhead{ID} & \colhead{(deg)} & \colhead{(deg)} & \colhead{(mag)} & \colhead{(mag)} & \colhead{(mag)} & \colhead{(mag)} & \colhead{color}} 

\startdata

501 & 53.1677662 & $-$27.8166951 & 24.73 & 24.47 & 23.94 & 23.40 & 1.068 \\
521 & 53.1699372 & $-$27.8178736 & 26.27 & 25.98 & 25.28 & 24.94 & 1.041 \\
901 & 53.1681788 & $-$27.8129432 & 24.44 & 24.03 & 23.35 & 23.06 & 0.970 \\
3048 & 53.1659121 & $-$27.7997159 & 26.63 & 26.21 & 25.40 & 25.17 & 1.041 \\
3299 & 53.1923333 & $-$27.7978741 & 25.71 & 25.40 & 25.00 & 24.59 & 0.817 \\
3372 & 53.1761793 & $-$27.7961337 & 22.90 & 22.34 & 21.64 & 21.25 & 1.095 \\
3373 & 53.1668679 & $-$27.7976989 & 24.48 & 24.43 & 23.94 & 23.77 & 0.662 \\
3613 & 53.1567527 & $-$27.7955981 & 23.80 & 23.34 & 22.63 & 22.12 & 1.219 \\
4072 & 53.1278092 & $-$27.7950828 & 24.66 & 24.49 & 24.26 & 23.73 & 0.759 \\
4084 & 53.1278372 & $-$27.7947976 & 24.61 & 24.44 & 24.05 & 23.56 & 0.888 \\
4438 & 53.1376781 & $-$27.7919373 & 23.58 & 23.13 & 22.44 & 22.08 & 1.049 \\
4491 & 53.1675702 & $-$27.7925214 & 23.93 & 23.68 & 23.24 & 22.89 & 0.791 \\
4591 & 53.1713278 & $-$27.7929428 & 25.08 & 24.83 & 24.16 & 23.84 & 0.990 \\
5190 & 53.1450905 & $-$27.7894219 & 24.22 & 24.00 & 23.70 & 23.17 & 0.827 \\
5405 & 53.1606000 & $-$27.7897302 & 26.09 & 25.79 & 25.12 & 24.64 & 1.150 \\
5417 & 53.1661791 & $-$27.7875215 & 23.10 & 22.61 & 21.97 & 21.48 & 1.135 \\
5658 & 53.1740361 & $-$27.7880062 & 24.99 & 24.61 & 23.99 & 23.43 & 1.178 \\
5805 & 53.1920649 & $-$27.7871824 & 23.91 & 23.58 & 23.04 & 22.57 & 1.011 \\
5989 & 53.1609549 & $-$27.7864996 & 24.81 & 24.58 & 24.05 & 23.55 & 1.028 \\
6079 & 53.1394080 & $-$27.7867760 & 25.76 & 25.35 & 24.78 & 24.12 & 1.222 \\
6785 & 53.1915603 & $-$27.7826687 & 24.15 & 23.88 & 23.57 & 22.97 & 0.907 \\
6821 & 53.1782201 & $-$27.7830771 & 24.13 & 23.90 & 23.47 & 23.04 & 0.865 \\
7036 & 53.1903447 & $-$27.7820005 & 24.43 & 24.29 & 24.09 & 23.57 & 0.714 \\
7112 & 53.1658806 & $-$27.7815379 & 24.65 & 24.09 & 23.41 & 22.83 & 1.258 \\
7559 & 53.1587381 & $-$27.7705348 & 23.93 & 23.54 & 22.86 & 22.48 & 1.057 \\
8125 & 53.1729989 & $-$27.7778393 & 23.95 & 23.85 & 23.45 & 23.14 & 0.716 \\
8551 & 53.1517635 & $-$27.7754183 & 23.90 & 23.42 & 22.71 & 22.25 & 1.173 \\
8585 & 53.1479310 & $-$27.7739569 & 22.42 & 22.13 & 21.63 & 21.15 & 0.973 \\
9018 & 53.1470806 & $-$27.7784243 & 24.44 & 24.32 & 23.96 & 23.66 & 0.663 \\
9125 & 53.1663274 & $-$27.7685923 & 24.13 & 23.83 & 23.36 & 22.71 & 1.121 \\
9183 & 53.1601763 & $-$27.7693039 & 24.86 & 24.80 & 24.40 & 24.18 & 0.611 \\
9341 & 53.1598624 & $-$27.7668404 & 24.88 & 24.68 & 24.12 & 23.72 & 0.953 \\
9444 & 53.1554162 & $-$27.7660748 & 25.81 & 24.91 & 24.04 & 23.34 & 1.572 \\
9759 & 53.1596949 & $-$27.7622834 & 24.76 & 24.62 & 24.34 & 23.82 & 0.803 \\

\enddata


\tablenotetext{$\dagger$}{Observed (\vv--\zz) color corresponds to rest-frame (\uu--\bb) color at \zgal.}
\tablecomments{Magnitudes are in the AB photometric system.}


\end{deluxetable}

\clearpage 


\begin{deluxetable}{cccccc}


\tabletypesize{\scriptsize}


\tablewidth{0pt}


\tablecaption{Our Sample - IDs, Coordinates and Redshifts\label{table2}}

\tablenum{2}

\tablehead{\colhead{HUDF} & \colhead{RA (J2000)} & \colhead{DEC (J2000)} & \colhead{$z^a$} & \colhead{$z^b$} & \colhead{$z^c$} \\ 
\colhead{(ID)} & \colhead{(deg)} & \colhead{(deg)} & \colhead{phot} & \colhead{VLT} & \colhead{SED Fit} } 

\startdata
  501  & 53.1677662  & $-$27.8166951 &   1.07 & \nodata &  1.02 \\
  521  & 53.1699372  & $-$27.8178736 &   1.04 & \nodata & 0.99 \\
  901  & 53.1681788  & $-$27.8129432 &   0.94 & \nodata & 0.88 \\
 3048  & 53.1659121  & $-$27.7997159 &   0.84 & \nodata & 0.85 \\
 3299  & 53.1923333  & $-$27.7978741 &   1.06 & 1.221   & 1.20 \\
 3372  & 53.1761793  & $-$27.7961337 &   1.04 & 0.996   & 0.98 \\
 3373  & 53.1668679  & $-$27.7976989 &   0.89 & \nodata & 0.93 \\
 3613  & 53.1567527  & $-$27.7955981 &   1.07 & 1.097   & 0.98 \\
 4072  & 53.1278092  & $-$27.7950828 &   1.27 & 1.189$^d$ & 1.20 \\
 4084  & 53.1278372  & $-$27.7947976 &   1.08 & \nodata & 1.06 \\
 4438  & 53.1376781  & $-$27.7919373 &   1.04 & 0.998   & 0.97 \\
 4491  & 53.1675702  & $-$27.7925214 &   1.05 & \nodata & 1.01 \\
 4591  & 53.1713278  & $-$27.7929428 &   0.92 & \nodata & 0.90 \\
 5190  & 53.1450905  & $-$27.7894219 &   1.22 & 1.316   & 1.18 \\
 5405  & 53.1606000  & $-$27.7897302 &   1.07 & 1.096   & 0.90 \\
 5417  & 53.1661791  & $-$27.7875215 &   1.10 & 1.097   & 1.03 \\
 5658  & 53.1740361  & $-$27.7880062 &   1.07 & 1.096   & 1.05 \\
 5805  & 53.1920649  & $-$27.7871824 &   1.05 & \nodata & 1.02 \\
 5989  & 53.1609549  & $-$27.7864996 &   0.95 & 1.135   & 1.00 \\
 6079  & 53.1394080  & $-$27.7867760 &   1.16 & 1.298   & 1.16 \\
 6785  & 53.1915603  & $-$27.7826687 &   1.14 & \nodata & 1.29 \\
 6821  & 53.1782201  & $-$27.7830771 &   1.07 & \nodata & 1.02 \\
 7036  & 53.1903447  & $-$27.7820005 &   1.21 & 0.743$^d$ & 1.16 \\
 7112  & 53.1658806  & $-$27.7815379 &   1.07 & \nodata & 1.04 \\
 7559  & 53.1587381  & $-$27.7705348 &   0.96 & \nodata & 0.93 \\
 8125  & 53.1729989  & $-$27.7778393 &   1.05 & \nodata & 1.04 \\
 8551  & 53.1517635  & $-$27.7754183 &   1.05$^c$ & 1.047 & 1.05 \\
 8585  & 53.1479310  & $-$27.7739569 &   0.97 & 1.088    & 1.05 \\ 
 9018  & 53.1470806  & $-$27.7784243 &   0.95 & \nodata  & 0.99 \\
 9125  & 53.1663274  & $-$27.7685923 &   1.20 & 1.295    & 1.21 \\
 9183  & 53.1601763  & $-$27.7693039 &   0.94 & \nodata  & 0.94 \\
 9341  & 53.1598624  & $-$27.7668404 &   1.03 & \nodata  & 1.01 \\
 9444  & 53.1554162  & $-$27.7660748 &   1.07 & 1.096    & 1.02 \\
 9759  & 53.1596949  & $-$27.7622834 &   1.34 & \nodata  & 1.22 \\
\enddata

\tablenotetext{a}{From GRAPES photometric redshift catalog \citep{ryan07}.}
\tablenotetext{b}{From GOODS-MUSIC catalog \citep{graz06} and \citet{vanz08}.} 
\tablenotetext{c}{From GRAPES SED fitting (\secref{models}), which is dominated by the 4000 \AA\ break fitting.}
\tablenotetext{d}{Quality flag on these spectroscopic redshifts is poor.}

\end{deluxetable}


\end{document}